\documentclass[twocolumn,journal]{IEEEtran}
\IEEEoverridecommandlockouts

\usepackage{color,graphicx,amsmath,amssymb,algorithm,algorithmic,amsmath,multirow,booktabs,array,amsthm,stfloats,caption,subfigure,bm,setspace,gensymb,url}

\newcommand{\be}{\begin{equation}}
\newcommand{\ee}{\end{equation}}
\newcommand{\bea}{\begin{eqnarray}}
\newcommand{\eea}{\end{eqnarray}}
\newcommand{\ba}{\begin{array}}
\newcommand{\ea}{\end{array}}
\newcommand{\nid}{\noindent}
\newcommand{\non}{\nonumber}

\renewcommand{\thesubsubsection}{\thesubsection.\arabic{subsubsection}}

\allowdisplaybreaks[4]

\title{Joint Transmit Waveform and Passive Beamforming Design for RIS-Aided DFRC Systems
\thanks{R. Liu, M. Li, and Y. Liu are with the School of Information and Communication Engineering, Dalian University of Technology, Dalian 116024, China (e-mail: liurang@mail.dlut.edu.cn; mli@dlut.edu.cn; yangliu{\_}613@dlut.edu.cn).}
\thanks{Q. Wu is with the State Key Laboratory of Internet of Things for Smart City, University of Macau, Macau 999078, China (email: qingqingwu@um.edu.mo).}
\thanks{Q. Liu is with the School of Computer Science and Technology, Dalian University of Technology, Dalian 116024, China (e-mail: qianliu@dlut.edu.cn).}
}

\author{Rang Liu,~\IEEEmembership{Graduate Student Member,~IEEE,}
        Ming Li,~\IEEEmembership{Senior Member,~IEEE,}
        Yang Liu,~\IEEEmembership{Member,~IEEE,}\\
        Qingqing Wu,~\IEEEmembership{Senior Member,~IEEE,}
        and Qian Liu,~\IEEEmembership{Member,~IEEE}
        }

\begin{document}

\maketitle

\pagestyle{empty}
\thispagestyle{empty}

\begin{abstract}
Reconfigurable intelligent surface (RIS) is a promising technology for 6G networks owing to its superior ability to enhance the capacity and coverage of wireless communications by smartly creating a favorable propagation environment. In this paper, we investigate the potential of employing RIS in dual-functional radar-communication (DFRC) systems for improving both radar sensing and communication functionalities. In particular, we consider a RIS-assisted DFRC system in which the multi-antenna base station (BS) simultaneously performs both multi-input multi-output (MIMO) radar sensing and multi-user multi-input single-output (MU-MISO) communications using the same hardware platform. We aim to jointly design the dual-functional transmit waveform and the passive beamforming of RIS to maximize the radar output signal-to-interference-plus-noise ratio (SINR) achieved by space-time adaptive processing (STAP), while satisfying the communication quality-of-service (QoS) requirement under one of three metrics, the constant-modulus constraint on the transmit waveform, and the unit-modulus constraint of RIS reflecting coefficients. An efficient algorithm framework based on the alternative direction method of multipliers (ADMM) and majorization-minimization (MM) methods is developed to solve the complicated non-convex optimization problem. Simulation results verify the advancement of the proposed RIS-assisted DRFC scheme and the effectiveness of the developed ADMM-MM-based joint transmit waveform and passive beamforming design algorithm.
\end{abstract}

\section{Introduction}\label{sec:introduction}

The recently commercialized fifth-generation (5G) wireless communication networks have achieved many technical improvements, including high data rate, massive connectivity, and low latency, enabled by several key technologies, such as massive multi-input multi-output (MIMO), millimeter-wave (mmWave) communications, and ultra-dense networking. Although 5G is still at an early stage of deployment, researchers are already seeking potential solutions for future sixth-generation (6G) system, which is envisioned as an integrated communication, sensing, and computing platform that enables various services \cite{Letaief CM 2019, Saad IEEE network 2020}.
Therefore, revolutionary technologies and radical changes are expected to pave the way for 6G networks to realize transformative encompassing applications.

Recent research reveals that reconfigurable intelligent surface (RIS) is a promising technology owing to its superior ability of boosting the capacity and coverage of wireless networks by smartly manipulating radio environment \cite{Renzo IWCN 2019}-\cite{Pan CM 2021}.
RIS, also called intelligent reflecting surface (IRS) \cite{Wu TCOM 2021} or large intelligent surface (LIS) \cite{Basar EuCNC 2019}, is a kind of meta-surface consisting of an array of passive reflecting elements, each of which can adjust the phase-shift, amplitude, frequency and polarization of the the incident signals.
By cooperatively controlling the parameters of the reflected electromagnetic (EM) waves that impinge on the elements of surface, RIS can generate passive reflection beamforming and create a more favorable propagation environment to greatly expand coverage, improve transmission quality, and enhance security, etc.
Furthermore, the employment of RIS can offer additional degrees of freedom (DoFs) for resolving severe channel fading and refining channel statistics.
Attracted by above advantages, RIS is deemed a crucial enabling technology for achieving intelligent radio environment in 6G networks and has received significant attentions from both industry and academia \cite{Pan JSTSP 2022}.

Motivated by the above, the applications of RIS in various wireless communication scenarios have been widely studied, including but not limited to the designs for maximizing spectral efficiency \cite{Yu ICCC 2019, Zhou WCL 2020}, sum-rate \cite{Pan TWC 2020}-\cite{Li TCOM 2021}, energy efficiency \cite{Huang TWC 2019, Yang TWC 2022}, and minimizing transmit power \cite{QWu TWC 2019, QWu TCOM 2020}, etc.
Since the ability of changing radio environments introduces a new optimization dimension, RIS is also considered in wireless systems in conjunction with other emerging techniques, such as simultaneous wireless information and power transfer (SWIPT) \cite{QWu JSAC 2020, Pan JSAC 2020}, unmanned aerial vehicle (UAV)-aided communications \cite{Hua TWC 2021}, physical-layer security \cite{Yu JSAC 2020}, and symbol-level precoding \cite{Rang TWC 2021, Rang TVT 2021}.
In addition, channel state information (CSI) acquisition, which is a crucial and challenging task in RIS-aided wireless systems, has been intensively investigated and comprehensively reviewed in \cite{Wang TWC 2020}-\cite{Swindlehurst 2021}.
With imperfect CSI, robust designs for RIS-aided communication systems were also studied \cite{Zhou TSP2 2020, YLiu TCOM 2021}.

On the other hand, it is foreseeable that the potential applications for 6G will demand not only high-quality ubiquitous wireless connectivity, but also high-accuracy and robust sensing capability.
Thus, there is a consensus that future 6G networks should provide both communication and sensing services, which motivates the recent research theme of integrated sensing and communications (ISAC) \cite{Zheng SPM 2019}-\cite{Zhang ICST 2022}.
The rationale of ISAC is that sensing and communication functions are tightly integrated and can be simultaneously performed using a fully-shared hardware platform, the same frequency bands, and a unified dual-functional waveform, which allows for significant improvements in hardware/spectrum/energy efficiency.
An ISAC system with such the tightest integration is more often referred to as a dual-functional radar-communication (DFRC) system in the literature.
The major research effort on DFRC systems is the design of novel dual-functional waveforms to simultaneously realize radar sensing and communication functions.
In recent years, MIMO architectures, which can provide spatial-domain DoFs, have been widely employed in DFRC systems \cite{FLiu TWC 2018}-\cite{Chen JSAC} to improve the waveform diversity for radar sensing, as well as to achieve beamforming gains and spatial multiplexing for multi-user communications.
Furthermore, space-time adaptive processing (STAP) technique \cite{Guerci 2014}-\cite{Tang TSP 2020}, which can offer additional temporal-domain DoFs, is utilized for MIMO-DFRC systems to achieve adaptive clutter suppression and better target detection by joint spatial-temporal waveform optimization \cite{RLiu JSAC 2022}.
After exploiting DoFs in both spatial and temporal domains, the propagation environment, which can be manipulated by RIS, is a potential new optimization dimension for DFRC systems to further enhance the performance of both functions.

While some works focused on exploiting RIS to enhance the sensing performance for radar systems, recent works \cite{Wang TVT 2021a}-\cite{Rang EUSIPCO2022} start investigating the integration of RIS for DFRC systems.
Specifically, RIS was deployed in DFRC systems for the purpose of mitigating multi-user interference (MUI) while assuring radar sensing performance in terms of transmit beampattern \cite{Wang TVT 2021a} or Cram\'{e}r-Rao bound \cite{Wang TVT 2021b}.
Although these works confirmed the advancement of deploying RIS in DFRC systems, the potential of RIS has not been fully exploited, since they only consider the impact of RIS on communication users regardless of its influence on the target.
Moreover, it is noted that the radar sensing functionality generally requires much higher transmit power, which leads to very strong received signals at the communication users.
Thus, it may be unwise to simply minimize the difference between the received signal and the desired symbol and irrationally eliminate all MUI, which will significantly restrict the flexibility of the transmit waveform design and deteriorate radar sensing performance.
In \cite{Jiang SJ 2021}, the signal-to-noise ratio (SNR) metric is utilized to evaluate the quality-of-service (QoS) of both radar sensing and communication.
Nevertheless, this initial work assumed a simplified and ideal system model, in which the base station (BS) serves only one user, and all possible paths that the transmit waveform can propagate through are not fully considered.
In very recent research \cite{Rang EUSIPCO2022}, the linear transmit beamforming and RIS reflection are jointly designed to maximize the sum-rate of communication users subject to the constraints of radar SNR, transmit power, and reflecting coefficients.
It is noted that the widespread clutter, which usually accompanies the target return, has not been considered in these existing works.

Motivated by above discussions, in this paper we attempt to fully exploit the potential of employing RIS in DFRC systems for enhancing radar sensing and communication functionalities.
In particular, we consider a multi-antenna BS that performs active sensing to detect a target in the presence of strong clutter and simultaneously transfers information symbols to multiple single-antenna users.
Our goal is to jointly design the transmit waveform and receive filter of the BS, and the reflecting coefficients of the RIS to improve the radar detection performance while assuring the QoS of multi-user multi-input single-output (MU-MISO) communications.
The main contributions of this paper can be summarized as follows:

\begin{itemize}
  \item We integrate a RIS in a DFRC system to achieve potential performance improvement by leveraging RIS capability of manipulating radio environment. For the considered novel RIS-assisted DFRC system, we present comprehensive signal models of radar and communication functionalities, and then introduce three practical metrics for evaluating the QoS of MU-MISO communications.

  \item We aim to jointly design the transmit waveform, the receive filter, and the passive beamforming (i.e., the reflecting coefficients of the RIS) to maximize the radar output signal-to-interference-plus-noise ratio (SINR), while satisfying the communication QoS requirement under any one of three metrics, the constant-modulus transmit waveform constraint, and the unit-modulus constraint of RIS reflecting coefficients. To solve the resulting complicated non-convex problem, the alternative direction method of multipliers (ADMM) and majorization-minimization (MM) methods are employed to convert it into several more tractable sub-problems, and then efficient algorithms are developed to alteratively solve them.

  \item Simulation studies demonstrate that the deployment of RIS can offer significant radar performance improvement especially for the DFRC systems with a weak direct channel between the BS and the target. Moreover, the proposed design algorithm can achieve satisfactory QoS for communications at the price of about 0.5dB radar performance loss compared with the radar-only system.
\end{itemize}

\emph{Notations}: Boldface lower-case and upper-case letters indicate column vectors and matrices, respectively.
$(\cdot)^*$, $(\cdot)^T$, $(\cdot)^H$, and $(\cdot)^{-1}$  denote  the conjugate, transpose, transpose-conjugate, and inverse operations, respectively.
$\mathbb{C}$ and $\mathbb{R}$ denote the sets of complex numbers and real numbers, respectively.
$| a |$, $\| \mathbf{a} \|$, and $\| \mathbf{A} \|_F$ are the magnitude of a scalar $a$, the norm of a vector $\mathbf{a}$, and the Frobenius norm of a matrix $\mathbf{A}$, respectively.
$\mathbb{E}\{\cdot\}$ represents statistical expectation.
$\text{Tr}\{\mathbf{A}\}$ takes the trace of the matrix $\mathbf{A}$ and $\text{vec}\{\mathbf{A}\}$ vectorizes the matrix $\mathbf{A}$.
$\otimes$ denotes the Kronecker product.
$\mathfrak{R}\{\cdot\}$ and $\mathfrak{I}\{\cdot\}$ denote the real and imaginary part of a complex number, respectively.
$\angle{a}$ is the angle of complex-valued $a$.
$\mathbf{I}_M$ indicates an $M\times M$ identity matrix.

\begin{figure}[!t]
\centering
\includegraphics[width = 3.2 in]{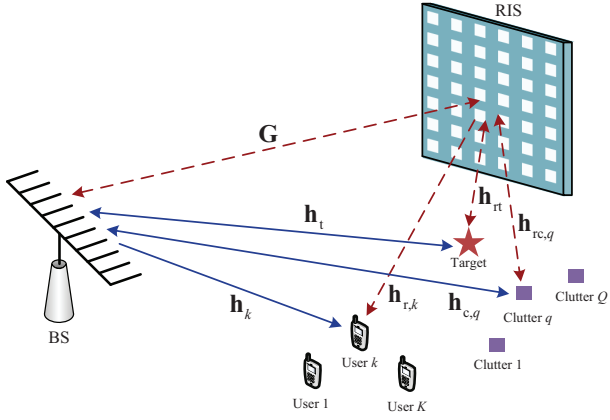}
\caption{A RIS-aided ISAC system.}\label{fig:system_model}\vspace{-0.1 cm}
\end{figure}

\section{System Model and Problem Formulation}\label{sec:system model}

We consider a RIS-aided colocated narrowband DFRC system as shown in Fig. 1, where a BS is equipped with $M$ transmit/receive antennas arranged as uniform linear arrays (ULAs).
The BS simultaneously performs radar and communication functionalities with the aid of an $N$-element RIS.
Specifically, the RIS assists the BS to detect a target in the presence of $Q$ clutter returns, and simultaneously transfer different symbols to $K$ single-antenna users.
With advanced self-interference mitigation techniques \cite{Sabharwal JSAC 2014}, we assume that the transmit and receive antennas of the BS work at the same time with perfect self-interference mitigation.
In order to achieve satisfactory target detection performance in the presence of strong signal-dependent clutter over widely spread ranges and angular regions, the BS utilizes the STAP technique to exploit all available DoFs in both spatial and temporal domains.
In other words, the BS uses STAP to optimize the nonlinear spatial-temporal transmit waveform and receive filter for offering better target detection performance.  The readers may refer to \cite{Guerci 2014}-\cite{Tang TSP 2020} for more details about the STAP technique.
Meanwhile, different information symbols are carried by the same transmit waveform, which is designed to manipulate the multiuser interference in a symbol-level manner for improving the communication QoS.

\subsection{Radar Model}

It is assumed that each radar pulse has $L$ digital samples and thus the range domain is divided into $L$ discrete bins.
Suppose that the target of interest locates in the range-angle position $(r_0,\theta_0)$ and the clutter sources locate in $(r_q,\theta_q),~r_q\in\{0,1,\ldots,L\},~q = 1, \ldots, Q$.
In this paper, we assume that the BS has the knowledge of the range-angle positions and the mean power of the clutter sources, which can be obtained from a cognitive paradigm \cite{Cognitive radar, Aubry TAES 2013} by using an environmental dynamic database including a geographical information system, digital terrain maps,clutter models, previous scanning/tracking files, etc.
For simplicity, we set the origin of the range coordinates as the target range, i.e., $r_0 = 0$.

Let $\mathbf{x}[l]\in\mathbb{C}^{M},~l = 1, \ldots, L$, be the $l$-th sample of the transmit waveform from the BS, and let $\bm{\phi} \triangleq [\phi_1,\ldots,\phi_N]^T$ with $|\phi_n| = 1,~n = 1,\ldots,N$, be the reflecting coefficients of the RIS.
As shown in Fig. 1, the transmit waveform will reach the target via the direct channel (the blue solid line) and the reflected channels (the red dashed lines), and then be reflected back to the BS through all the propagation paths.
Thus, the target echo signal can be obtained by propagating the transmit waveform through four different paths, i.e., BS-target-BS, BS-RIS-target-BS, BS-target-RIS-BS, and BS-RIS-target-RIS-BS.
We assume that the differences of the propagation delays between these four paths are negligible, considering that the distance between the BS and the RIS is usually much larger than the distances between the RIS and the target/clutter sources.
Therefore, the received baseband signal at the BS in the $l$-th time slot, which consists of the target echo signal, the clutter returns and the noise, can be written as

\small\vspace{-0.3 cm}
\begin{equation}\label{eq:rl}
\mathbf{r}[l] =\alpha_0(\mathbf{h}_\text{t}
+\mathbf{G}^H\bm{\Phi}\mathbf{h}_\text{rt})(\mathbf{h}^H_\text{t}+\mathbf{h}^H_\text{rt}\bm{\Phi}\mathbf{G})
\mathbf{x}[l]e^{\jmath2\pi(l-1)\nu_0} + \mathbf{c}[l] + \mathbf{z}[l].
\end{equation}
\normalsize In (\ref{eq:rl}), the vectors $\mathbf{h}_\text{t}\in\mathbb{C}^M$ and $\mathbf{h}_\text{rt}\in\mathbb{C}^N$ respectively denote the channels between the BS and the target and between the RIS and the target, which are generally line-of-sight (LoS) in the field of radar and are determined by the range-angle position $(r_0,\theta_0)$.
The matrix $\mathbf{G}\in\mathbb{C}^{N\times M}$ represents the channel between the BS and the RIS.
The RIS reflection matrix is defined by $\bm{\Phi} \triangleq \text{diag}\{\bm{\phi}\}$.
The scalar $\alpha_0$ denotes the target radar cross section (RCS) and $\mathbb{E}\{|\alpha_0|^2\} = \varsigma_0^2$.
The scalar $\nu_0$ is the Doppler frequency of the target.
The overall clutter return $\mathbf{c}[l]$, which follows the similar propagation paths as the target echo signal, can be expressed as

\small\vspace{-0.3 cm}
\begin{equation}\mathbf{c}[l] =\hspace{-0.05cm}\sum_{q=1}^Q\alpha_q(\mathbf{h}_{\text{c},q}+\mathbf{G}^H\bm{\Phi}\mathbf{h}_{\text{rc},q})
(\mathbf{h}_{\text{c},q}^H+\mathbf{h}_{\text{rc},q}^H\bm{\Phi}\mathbf{G})
\mathbf{x}[l-r_q]e^{\jmath2\pi(l-1)\nu_q},
\end{equation}
\normalsize where the scalar $\alpha_q$ denotes the complex amplitude of the $q$-th clutter, $\mathbb{E}\{|\alpha_q|^2\} = \varsigma_q^2$. The vectors $\mathbf{h}_{\text{c},q}\in\mathbb{C}^M$ and $\mathbf{h}_{\text{rc},q}\in\mathbb{C}^N$ represent the channels between the BS and the $q$-th clutter source and between the RIS and the $q$-th clutter source, respectively.
The vector $\mathbf{z}[l]\in\mathbb{C}^M$ is additive white Gaussian noise (AWGN) and $\mathbf{z}[l]\sim \mathcal{CN}(\mathbf{0},\varsigma_\text{z}^2\mathbf{I}_M)$.
For simplicity, in this paper we assume both the target and clutter sources are slowly-moving objects, whose Doppler frequencies equal to zeros, i.e., $\nu_0 = \nu_q = 0,~\forall q$.

For conciseness, we define
\begin{subequations}
\begin{align}
\mathbf{H}_0(\bm{\phi}) &\triangleq (\mathbf{h}_\text{t}
+\mathbf{G}^H\bm{\Phi}\mathbf{h}_\text{rt})(\mathbf{h}^H_\text{t}+\mathbf{h}^H_\text{rt}\bm{\Phi}\mathbf{G}),\\
\mathbf{H}_q(\bm{\phi}) &\triangleq (\mathbf{h}_{\text{c},q}+\mathbf{G}^H\bm{\Phi}\mathbf{h}_{\text{rc},q})
(\mathbf{h}_{\text{c},q}^H+\mathbf{h}_{\text{rc},q}^H\bm{\Phi}\mathbf{G}),
\end{align}
\end{subequations}
which can be intuitively seen as the effective channels for the target return and the $q$-th clutter return, respectively.
Since the RIS reflection matrix $\bm{\Phi}$ is diagonal, we have $\mathbf{h}^H_\text{rt}\bm{\Phi} = \bm{\phi}^T\text{diag}\{\mathbf{h}^H_\text{rt}\}$.
Thus, by defining
\begin{equation}\label{eq:AB def}
\begin{aligned}
\mathbf{A}_0 &\triangleq \mathbf{h}_\text{t}\mathbf{h}_\text{t}^H, \qquad\qquad\quad  \mathbf{B}_0 \triangleq \text{diag}\{\mathbf{h}^H_\text{rt}\}\mathbf{G},\\
\mathbf{A}_q &\triangleq \mathbf{h}_{\text{c},q}\mathbf{h}_{\text{c},q}^H, \qquad\qquad \mathbf{B}_q \triangleq \text{diag}\{\mathbf{h}_{\text{rc},q}^H\}\mathbf{G},
\end{aligned}
\end{equation}
the effective channels $\mathbf{H}_0(\bm{\phi})$ and $\mathbf{H}_q(\bm{\phi})$ can be concisely re-arranged as
\begin{subequations}\label{eq:fq}
\begin{align}
\mathbf{H}_0(\bm{\phi}) &= \mathbf{A}_0 + \mathbf{h}_\text{t}\bm{\phi}^T\mathbf{B}_0
+\mathbf{B}_0^H\bm{\phi}\mathbf{h}_\text{t}^H+\mathbf{B}_0^H\bm{\phi}\bm{\phi}^T\mathbf{B}_0,\\
\mathbf{H}_q(\bm{\phi}) &= \mathbf{A}_q +\mathbf{h}_{\text{c},q}\bm{\phi}^T\mathbf{B}_q
+\mathbf{B}_q^H\bm{\phi}\mathbf{h}_{\text{c},q}^H+\mathbf{B}_q^H\bm{\phi}\bm{\phi}^T\mathbf{B}_q.
\end{align}
\end{subequations}
With the expressions (\ref{eq:fq}), the received signal in the $l$-th time slot $\mathbf{r}[l]$ can be re-written as
\begin{equation}
\mathbf{r}[l] = \alpha_0\mathbf{H}_0(\bm{\phi})\mathbf{x}[l] + \sum_{q=1}^Q\alpha_q\mathbf{H}_q(\bm{\phi})\mathbf{x}[l-r_q] + \mathbf{z}[l].
\end{equation}

Then, we stack the received signals during a radar pulse by defining $\mathbf{r}\triangleq[\mathbf{r}[1]^T,\ldots,\mathbf{r}[L]^T]^T$, $\mathbf{x}\triangleq[\mathbf{x}[1]^T,\ldots,\mathbf{x}[L]^T]^T$, and $\mathbf{z}\triangleq[\mathbf{z}[1]^T,\ldots,\mathbf{z}[L]^T]^T$.
The received signals of $L$ samples can thus be expressed as
\begin{equation}
\mathbf{r} = \alpha_0\widetilde{\mathbf{H}}_0(\bm{\phi})\mathbf{x} + \sum_{q=1}^Q\alpha_q\widetilde{\mathbf{H}}_q(\bm{\phi})\mathbf{x} + \mathbf{z},
\end{equation}
where for simplicity we define
\begin{equation}\label{eq:fqtilde}
\widetilde{\mathbf{H}}_0(\bm{\phi}) \triangleq \mathbf{I}_L\otimes\mathbf{H}_0(\bm{\phi}),\quad~~\widetilde{\mathbf{H}}_q(\bm{\phi}) \triangleq [\mathbf{I}_L\otimes\mathbf{H}_q(\bm{\phi})]\mathbf{J}_{r_q}.
\end{equation}
The shift matrix $\mathbf{J}_{r_q}\in\mathbb{R}^{ML\times ML}$, which represents the propagation delay of the clutter returns from different ranges, is defined by
$
\mathbf{J}_{r_q}(i,j) = \left\{
             \begin{array}{lr}1,~~~i-j = Mr_q,&\\
             0, ~~~\text{otherwise}.&
             \end{array}
\right.
$

The received spatial-temporal echo signals are processed through an associated linear space-time receive filter $\mathbf{w}\in\mathbb{C}^{ML}$, and then the radar output at the BS is obtained as
\begin{equation}
\mathbf{w}^H\mathbf{r} = \alpha_0\mathbf{w}^H\widetilde{\mathbf{H}}_0(\bm{\phi})\mathbf{x} + \mathbf{w}^H\sum_{q=1}^Q\alpha_q\widetilde{\mathbf{H}}_q(\bm{\phi})\mathbf{x} + \mathbf{w}^H\mathbf{z}.
\end{equation}
Thus, the radar output SINR is given by
\begin{equation}\label{eq:gamma}
\gamma_\text{r} = \frac{\varsigma_0^2|\mathbf{w}^H\widetilde{\mathbf{H}}_0(\bm{\phi})\mathbf{x}|^2}
{\mathbf{w}^H\big[\sum_{q=1}^Q\varsigma_q^2\widetilde{\mathbf{H}}_q(\bm{\phi})\mathbf{xx}^H
\widetilde{\mathbf{H}}^H_q(\bm{\phi})+\varsigma_\text{z}^2\mathbf{I}_{ML}\big]\mathbf{w}}.
\end{equation}
Since the target detection probability monotonically increases with the radar output SINR $\gamma_\text{r}$, in this paper we aim to maximize $\gamma_\text{r}$ to achieve better target detection performance.

Considering that the echo signal from the target is usually very weak, sufficient transmit power is required at the BS to provide satisfactory target detection performance.
Potential large peak-to-average power ratio (PAPR) with this high power will cause significant nonlinear distortions to practical hardware circuits.
In the sequel, constant-modulus waveforms are usually desired in practical radar systems, i.e.,
\begin{equation}\label{eq:power constraint}
|x_i| = \sqrt{P/M},~~\forall i = 1,\ldots,ML,
\end{equation}
where $x_i$ denotes the $i$-th entry of $\mathbf{x}$ and $P$ is the available transmit power of the $M$ antennas.

\subsection{Communication Model}

In addition to detect the target of interest, the transmit waveform $\mathbf{x}$ simultaneously carries different information symbols for $K$ single-antenna users.
In particular, denoting the desired symbols of the $K$ users in the $l$-th time slot as $\mathbf{s}[l]\triangleq[s_1[l],\ldots,s_K[l]]^T$, where the information symbol $s_k[l]$ is independently selected from $\Omega$-phase-shift-keying (PSK) constellations.
The received signal at the $k$-th user can be expressed as
\begin{equation}
r_{k}[l] = (\mathbf{h}_{k}^H + \mathbf{h}_{\text{r},k}^H\bm{\Phi}\mathbf{G})\mathbf{x}[l] + n_{k}[l],~~\forall l,
\end{equation}
where the vectors $\mathbf{h}_k\in\mathbb{C}^M$ and $\mathbf{h}_{\text{r},k}\in\mathbb{C}^N$ represent the channels between the BS and the $k$-th user and between the RIS and the $k$-th user, respectively\footnote{Given that various advanced algorithms have been proposed in \cite{Wang TWC 2020}-\cite{Swindlehurst 2021} and the references therein to handle the CSI acquisition problem for RIS-assisted systems, we assume perfect CSI in this paper in order to focus on exploring the potential of RIS in DFRC system. Problems involving imperfect CSI or no CSI are left for future studies.}.
The scalar $n_k[l]$ is AWGN to the $k$-th user and $n_k[l]\sim\mathcal{CN}(0,\sigma_k^2)$.
Given preset constellation information, the BS should elaborately design the transmit waveform signal $\mathbf{x}[l]$ such that the receivers can easily decode the desired symbol $s_k[l]$ from the received noise-corrupted signal $r_k[l]$.

In order to evaluate the communication QoS, we introduce three typical metrics according to their different strategies of handling MUI: \textit{i}) zero-forcing (ZF)-type, \textit{ii}) minimum mean square error (MMSE)-type, and \textit{iii}) constructive interference (CI)-type.
Without loss of generality, we take the quadrature-PSK (QPSK) modulation as an example.
We assume that the desired symbol of the $k$-th user in the $l$-th time slot is $s_k[l] = (1/\sqrt{2},\jmath/\sqrt{2})$ and show the received noise-free signals in the complex plane as in Fig. 2, where $\theta$ represents half of the angular range of the decision region, $\theta = \pi/\Omega$.
Point $D$ denotes the received noise-free signal $\widetilde{r}_k[l]$,
\be
\overrightarrow{OD} = \widetilde{r}_k[l] = (\mathbf{h}_{k}^H + \mathbf{h}_{\text{r},k}^H\bm{\Phi}\mathbf{G})\mathbf{x}[l].
\ee
To quantify the communication QoS, we let $\Gamma_k$ denote the required SNR of the $k$-th user.
Point $A$ represents the desired symbol with the required SNR, i.e., $\overrightarrow{OA} =\sigma_k\sqrt{\Gamma_{k}}s_{k}[l]$.
Thus, the MUI can be seen as the difference between the received noise-free signal and the desired symbol with the required SNR, i.e., $\overrightarrow{OD}-\overrightarrow{OA} = (\mathbf{h}_{k}^H + \mathbf{h}_{\text{r},k}^H\bm{\Phi}\mathbf{G})\mathbf{x}[l]-\sigma_k\sqrt{\Gamma_{k}}s_{k}[l]$.

\begin{figure}[!t]
\centering
\includegraphics[width = 2.1 in]{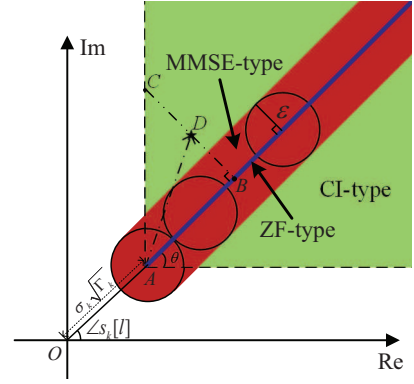}\vspace{0.1 cm}
\caption{Illustration of three metrics for communication QoS.}\label{fig:CR}\vspace{-0.2cm}
\end{figure}

1) ZF-type communication metric: Conventional ZF-based metrics aim to entirely eliminate the MUI by ensuring that the transmit signal $\mathbf{x}[l]$ satisfies
\be\label{eq:old zf}
(\mathbf{h}_{k}^H + \mathbf{h}_{\text{r},k}^H\bm{\Phi}\mathbf{G})\mathbf{x}[l]=\sigma_k\sqrt{\Gamma_{k}}s_{k}[l],
\ee
which enforces the received noise-free signal being located at point $A$.
In practical DFRC systems, however, the transmit power at the BS is usually very large for achieving a satisfactory radar output strength, which leads to a significantly stronger received signal than the required level $\sigma_k\sqrt{\Gamma_{k}}s_{k}[l]$.
Therefore, enforcing the received noise-free signal being located exactly at point $A$ is unwise.
Moreover, this original ZF constraint imposes too stringent restrictions onto the waveform design, which may seriously deteriorate the radar sensing functionality.
To tackle this issue, in this paper we propose a new rational ZF-type communication metric by imposing a scaling factor  $\alpha_{k,l} \geq 1, \forall k,l$, to the desired symbol and modifying (\ref{eq:old zf}) as
\begin{equation}\label{eq:ZF original constraint}
(\mathbf{h}_{k}^H \hspace{-0.05cm}+ \hspace{-0.05cm} \mathbf{h}_{\text{r},k}^H\bm{\Phi}\mathbf{G})\mathbf{x}[l]=\alpha_{k,l}\sigma_k\sqrt{\Gamma_{k}}s_{k}[l],~\alpha_{k,l} \geq 1,~\forall k,l,
\end{equation}
by which the MUI is tuned to the same direction as the desired symbol $s_{k}[l]$ and thus the received noise-free signals are located on the extension of $\overrightarrow{OA}$ (the blue ray in Fig. 2).
In the sequel, the QoS requirement is satisfied and more flexibility can be provided for the radar sensing functionality.

2) MMSE-type communication metric: The MMSE-type metric compromises the strict equality constraint in (\ref{eq:ZF original constraint}) by tolerating certain variations, i.e., the square error between the received noise-free signal $\widetilde{r}_k[l]$ and the scaled desired symbol $\alpha_{k,l}\sigma_k\sqrt{\Gamma_{k}}s_{k}[l]$, within a certain level:
\begin{equation}\label{eq:MMSE metric}
\big|(\mathbf{h}_{k}^H \hspace{-0.05cm}+ \hspace{-0.05cm} \mathbf{h}_{\text{r},k}^H\bm{\Phi}\mathbf{G})\mathbf{x}[l]\hspace{-0.05cm}-\hspace{-0.05cm}\alpha_{k,l}\sigma_k\sqrt{\Gamma_{k}}s_{k}[l] \big|^2 \hspace{-0.05cm}\leq\hspace{-0.05cm} \epsilon,~\alpha_{k,l} \geq 1,~\forall k,l,
\end{equation}
where $\epsilon$ is a preset threshold for assuring the MUI suppression.
With the MMSE-type metric (\ref{eq:MMSE metric}), the received noise-free signal will lie within a disk centered at $\alpha_{k,l}\sigma_k\sqrt{\Gamma_{k}}s_{k}[l]$, and all possible signals satisfying (\ref{eq:MMSE metric}) form a red region shown in Fig. 2, which is greatly larger than that with the ZF-type metric (\ref{eq:ZF original constraint}).
Thus, the MMSE-type metric can exploit more flexibility for improving the radar sensing functionality compared to the ZF-type metric.

3) CI-type communication metric: Instead of forcing the MUI to lie in the exact direction of the desired symbol $s_k[l]$ or restricting it within a certain region along that direction, the CI-type communication metric offers more available flexibility for handling MUI.
Borrowing the idea from symbol-level precoding \cite{Rang TWC 2021, Rang TVT 2021}, the MUI is constructive for the symbol detection if it can push the received noise-free signal away from its corresponding decision boundaries, for example the positive halves of the $x$ and $y$ axes in Fig. 2.
In other words, the communication QoS can be guaranteed when the MUI forces the received noise-free signal deeper into its corresponding constructive region (the green region in Fig. 2).
In order to explicitly express this constraint, we project point $D$ onto the direction of $\overrightarrow{OA}$ at point $B$, and extend $\overrightarrow{BD}$ to intersect with the nearest boundary of the constructive region at point $C$.
Then, the relationship guaranteeing that the received noise-free signal is located in the green constructive region is expressed as $|\overrightarrow{BC}|-|\overrightarrow{BD}|\geq 0$, which can be formulated as
\begin{equation}\begin{aligned}
&\Re\big\{(\mathbf{h}_{k}^H \hspace{-0.05 cm}+ \hspace{-0.05 cm} \mathbf{h}_{\text{r},k}^H\bm{\Phi}\mathbf{G})\mathbf{x}[l]e^{-\jmath\angle{s_{k}[l]}}
\hspace{-0.05 cm}-\hspace{-0.05 cm}\sigma_{k}\sqrt{\Gamma_{k}}\big\}\sin \theta \\
&~-\hspace{-0.05 cm}\big|\Im\big\{(\mathbf{h}_{k}^H\hspace{-0.08 cm} +\hspace{-0.08 cm} \mathbf{h}_{\text{r},k}^H\bm{\Phi}\mathbf{G})\mathbf{x}[l]e^{-\jmath\angle{s_{k}[l]}}\big\}\big|\hspace{-0.05 cm}\cos\theta
   \geq 0,~\forall k,l,
\end{aligned}\end{equation}
and further be equivalently re-written as
\be\label{eq:CI original constraint}
\Re\Big\{\big(\mathbf{h}_{k}^H+ \mathbf{h}_{\text{r},k}^H\bm{\Phi}\mathbf{G}\big)\mathbf{x}[l]\frac{e^{-\jmath\angle s_k[l]}
(\sin\theta\pm e^{-\jmath\pi/2}\cos\theta)}{\sigma_{k}\sqrt{\Gamma_{k}}\sin\theta}\Big\} \geq 1.
\ee
Due to space limitations, the detailed derivations about this constraint are omitted here.
The readers can refer to the literature about CI-based symbol-level precoding \cite{Rang TWC 2021, Rang TVT 2021}.

Based on the above discussions and formulations, it is clear that the ZF-type metric allows the least DoFs since it forces the MUI to be at the same direction as the desired symbol, while the CI-type metric allows the largest DoFs in optimizing the transmit waveform considering that it permits the MUI to be in the whole constructive region.
Therefore, for the same SNR requirement $\Gamma_k$, a trade-off between target detection and communication performance is expected in solving the resulting problems under these communication metrics, i.e., the solution obtained under the ZF-type metric achieves the worst target detection and the best communication performance, while the opposite result is achieved under the CI-type metric.

In order to explicitly express constraints (\ref{eq:ZF original constraint})-(\ref{eq:CI original constraint}) with respect to the transmit waveform $\mathbf{x}$ and the RIS reflecting coefficients $\bm{\phi}$, we stack the transmit signals $\mathbf{x}[l],~ \forall l$, and utilize the transformation $\mathbf{h}_{\text{r},k}^H\bm{\Phi} = \bm{\phi}^T\text{diag}\{\mathbf{h}_{\text{r},k}^H\}$ to recast the received noise-free signal $\widetilde{r}_k[l]$ as
\be
\widetilde{r}_k[l] = \left[\mathbf{e}_l^T \otimes \left(\mathbf{h}_{k}^H + \bm{\phi}^T \text{diag}\{\mathbf{h}_{\text{r},k}^H\}\mathbf{G}\right)\right]\mathbf{x},
\ee
where the vector $\mathbf{e}_l$ is the $l$-th column of the identity matrix $\mathbf{I}_L$.
Then, by defining
\begin{subequations}
\begin{align}
\mathbf{h}_{k,l} &\triangleq \mathbf{e}_l \otimes \mathbf{h}_k,\\
\mathbf{G}_{k} &\triangleq \mathbf{I}_L\otimes \text{diag}\{\mathbf{h}_{\text{r},k}^H\}\mathbf{G},\\
\gamma_{k,l}^\text{ZF} &\triangleq \sigma_k\sqrt{\Gamma_k}s_k[l],\\
\gamma_{k,l}^\text{CI} &\triangleq \frac{e^{-\jmath\angle s_k[l]}
(\sin\theta\pm e^{-\jmath\pi/2}\cos\theta)}{\sigma_k\sqrt{\Gamma_k}\sin\theta},
\end{align}
\end{subequations}
the ZF-type, MMSE-type, and CI-type communication QoS constraints can be respectively re-formulated as
\begin{subequations}\label{eq:communication QoS constraints}
\begin{align}
&\big[\mathbf{h}_{k,l}^H\hspace{-0.05cm}+\hspace{-0.05cm}(\mathbf{e}_l^T\otimes\bm{\phi}^T)\mathbf{G}_{k}\big]\mathbf{x} = \alpha_{k,l}\gamma_{k,l}^\text{ZF},~~\alpha_{k,l} \geq 1,~\forall k,l,\\
&\big|[\mathbf{h}_{k,l}^H\hspace{-0.07cm}+\hspace{-0.07cm}(\mathbf{e}_l^T\hspace{-0.08cm}\otimes\hspace{-0.08cm}\bm{\phi}^T\hspace{-0.05cm})
\mathbf{G}_{k}]\mathbf{x}\hspace{-0.07cm}-\hspace{-0.07cm}\alpha_{k,l}\gamma_{k,l}^\text{ZF}\big|^2\hspace{-0.1cm}\leq\hspace{-0.07cm} \epsilon,~\alpha_{k,l}\hspace{-0.07cm}\geq\hspace{-0.07cm}1,~\forall k,l,\\
&\Re\big\{\gamma_{k,l}^\text{CI}[\mathbf{h}^H_{k,l}+(\mathbf{e}_l^T\otimes\bm{\phi}^T)\mathbf{G}_{k}]\mathbf{x}\big\}
\geq 1,~~\forall k,l.
\end{align}
\end{subequations}

\subsection{Problem Formulation}

In this paper, we investigate to jointly design the transmit waveform $\mathbf{x}$, the scaling factor $\bm{\alpha} \triangleq [\alpha_{1,1}, \alpha_{1,2}, \ldots, \alpha_{K,L}]^T$, the receive filter $\mathbf{w}$, and the reflecting coefficients $\bm{\phi}$ to maximize the radar output SINR, while satisfying the communication QoS requirement under one of the metrics (\ref{eq:communication QoS constraints}), the constant-modulus waveform constraint (\ref{eq:power constraint}), and the unit-modulus constraint of RIS reflecting coefficients.
Therefore, the optimization problem is formulated as
\begin{subequations}\label{eq:original problem}
\begin{align}
&\underset{\mathbf{x},\bm{\alpha},\mathbf{w},\bm{\phi}}\max~~\frac{\varsigma_0^2|\mathbf{w}^H\widetilde{\mathbf{H}}_0(\bm{\phi})\mathbf{x}|^2}
{\mathbf{w}^H\big[\sum_{q=1}^Q\varsigma_q^2\widetilde{\mathbf{H}}_q(\bm{\phi})\mathbf{xx}^H
\widetilde{\mathbf{H}}^H_q(\bm{\phi})+\varsigma_\text{z}^2\mathbf{I}_{ML}\big]\mathbf{w}}\\
&\quad\text{s.t.}\quad(\ref{eq:communication QoS constraints}\text{a})~\text{or}~(\ref{eq:communication QoS constraints}\text{b})~\text{or}~ (\ref{eq:communication QoS constraints}\text{c}), \\
&\quad\quad\quad|x_i| = \sqrt{P/M},~~\forall i,\\
&\quad\quad\quad|\phi_n| = 1,~~\forall n.
\end{align}
\end{subequations}

It is easy to see that with given transmit waveform $\mathbf{x}$, scaling factor $\bm{\alpha}$, and RIS reflecting coefficients $\bm{\phi}$, the optimization for the receive filter $\mathbf{w}$ is reduced to a minimum variance distortionless response (MVDR) problem:
\begin{subequations}
\begin{align}
&\underset{\mathbf{w}}\min~~~\mathbf{w}^H\Big[\sum_{q=1}^Q\varsigma_q^2\widetilde{\mathbf{H}}_q(\bm{\phi})\mathbf{xx}^H\widetilde{\mathbf{H}}^H_q(\bm{\phi})
+\varsigma_\text{z}^2\mathbf{I}\Big]\mathbf{w}\\
&~\text{s.t.}~~~~\mathbf{w}^H\widetilde{\mathbf{H}}_0(\bm{\phi})\mathbf{x} = 1,
\end{align}
\end{subequations}
whose optimal solution can be easily obtained by
\begin{equation}\label{eq:optimal w}
\mathbf{w}^\star\hspace{-0.1cm} =\hspace{-0.1cm} \frac{\big[\sum_{q=1}^Q\varsigma_q^2\widetilde{\mathbf{H}}_q(\bm{\phi})\mathbf{xx}^H\widetilde{\mathbf{H}}^H_q(\bm{\phi})
+\varsigma_\text{z}^2\mathbf{I}\big]^{-1}\widetilde{\mathbf{H}}_0(\bm{\phi})\mathbf{x}}
{\mathbf{x}^H\widetilde{\mathbf{H}}^H_0(\hspace{-0.02cm}\bm{\phi}\hspace{-0.02cm})\big[\hspace{-0.1cm}\sum_{q=1}^Q\hspace{-0.1cm}\varsigma_q^2
\widetilde{\mathbf{H}}_q\hspace{-0.05cm}(\hspace{-0.02cm}\bm{\phi}\hspace{-0.02cm})\mathbf{xx}^H\widetilde{\mathbf{H}}^H_q(\hspace{-0.02cm}\bm{\phi}\hspace{-0.02cm})
\hspace{-0.1cm}+\hspace{-0.1cm}\varsigma_\text{z}^2\mathbf{I}\big]^{-1}\widetilde{\mathbf{H}}_0(\hspace{-0.02cm}\bm{\phi}\hspace{-0.02cm})\mathbf{x}}.
\end{equation}
Substituting the optimal receive filter (\ref{eq:optimal w}) into (\ref{eq:original problem}a), the original optimization problem is transformed into
\begin{subequations}\label{eq:problem2}
\begin{align}
&\underset{\mathbf{x},\bm{\alpha},\bm{\phi}}\min-\hspace{-0.05cm}\mathbf{x}^H\widetilde{\mathbf{H}}^H_0\hspace{-0.02cm}(\hspace{-0.02cm}\bm{\phi}\hspace{-0.02cm})
\hspace{-0.02cm}\Big[\hspace{-0.04cm}\sum_{q=1}^Q\hspace{-0.02cm}\varsigma_q^2\widetilde{\mathbf{H}}_q(\hspace{-0.02cm}\bm{\phi}\hspace{-0.02cm})
\mathbf{xx}^H\widetilde{\mathbf{H}}^H_q(\hspace{-0.02cm}\bm{\phi}\hspace{-0.02cm})
\hspace{-0.04cm}+\hspace{-0.04cm}\varsigma_\text{z}^2\mathbf{I}\Big]^{-1}\hspace{-0.02cm}\widetilde{\mathbf{H}}_0(\hspace{-0.02cm}\bm{\phi}\hspace{-0.02cm})\mathbf{x}\\
&~\text{s.t.}~~(\ref{eq:communication QoS constraints}\text{a})~\text{or}~(\ref{eq:communication QoS constraints}\text{b})~\text{or}~ (\ref{eq:communication QoS constraints}\text{c}), \\
&~~~~~~|x_i| = \sqrt{P/M},~~\forall i,\\
&~~~~~~|\phi_n| = 1,~~\forall n.
\end{align}
\end{subequations}

We observe that problem (\ref{eq:problem2}) is a complicated non-convex problem due to the non-convex bivariate objective (\ref{eq:problem2}a) and the constant-modulus constraints on both the transmit waveform (\ref{eq:problem2}c) and the reflecting coefficients (\ref{eq:problem2}d).
In order to tackle these difficulties, in the next section we propose to utilize the ADMM and MM methods to convert it into several more tractable sub-problems, and then develop efficient algorithms to alteratively solve them.

\section{Joint Transmit Waveform and Passive Beamforming Design}

\subsection{ADMM-based Transformation}

In order to handle the non-convex constant-modulus constraints (\ref{eq:problem2}c) and (\ref{eq:problem2}d), we first introduce two auxiliary vectors $\mathbf{y}\triangleq[y_1,\ldots,y_{ML}]^T$ and $\bm{\varphi}\triangleq[\varphi_1,\ldots,\varphi_N]^T$ to convert problem (\ref{eq:problem2}) into
\begin{subequations}\label{eq:problem3}
\begin{align}
&\underset{\mathbf{x},\bm{\alpha},\mathbf{y},\bm{\varphi},\bm{\phi}}\min~f_1(\mathbf{x},\bm{\phi})\\
&\quad~\text{s.t.}\quad~~(\ref{eq:communication QoS constraints}\text{a})~\text{or}~(\ref{eq:communication QoS constraints}\text{b})~\text{or}~ (\ref{eq:communication QoS constraints}\text{c}), \\
&\hspace{1.45 cm}|x_i| \leq \sqrt{P/M},~~\forall i,\\
&\hspace{1.45 cm}|\phi_n| \leq 1,~~\forall n,\\
&\hspace{1.45 cm}\mathbf{y} = \mathbf{x},\\
&\hspace{1.45 cm}\bm{\varphi} = \bm{\phi},\\
&\hspace{1.45 cm}|y_i| = \sqrt{P/M},~~\forall i,\\
&\hspace{1.45 cm}|\varphi_n| = 1,~~\forall n,
\end{align}
\end{subequations}
where for simplicity we define the objective function as
\be\begin{aligned}
&f_1(\mathbf{x},\bm{\phi}) \triangleq\\
& -\mathbf{x}^H\widetilde{\mathbf{H}}^H_0(\bm{\phi})
\big[\sum_{q=1}^Q\varsigma_q^2\widetilde{\mathbf{H}}_q(\bm{\phi})\mathbf{xx}^H\widetilde{\mathbf{H}}^H_q(\bm{\phi})
+\varsigma_\text{z}^2\mathbf{I}\big]^{-1}\widetilde{\mathbf{H}}_0(\bm{\phi})\mathbf{x}.
\end{aligned}\ee
To accommodate the ADMM framework and facilitate the algorithm development \cite{YLiu TCOM 2021}, we define an indicator function $\mathbb{I}(\mathbf{x},\bm{\alpha},\mathbf{y},\bm{\phi},\bm{\varphi})$ to impose constraints (\ref{eq:problem3}b)-(\ref{eq:problem3}d), (\ref{eq:problem3}g), and (\ref{eq:problem3}h) in the objective.
Specifically, the value of $\mathbb{I}(\mathbf{x},\bm{\alpha},\mathbf{y},\bm{\phi},\bm{\varphi})$ equals zero when the variables $\mathbf{x}$, $\bm{\alpha}$, $\mathbf{y}$, $\bm{\phi}$, and $\bm{\varphi}$ lie in the feasible regions endowed by constraints (\ref{eq:problem3}b)-(\ref{eq:problem3}d), (\ref{eq:problem3}g), and (\ref{eq:problem3}h); otherwise it goes to infinite.
Then, the optimization problem is transformed into
\begin{subequations}
\begin{align}
&\underset{\mathbf{x},\bm{\alpha},\mathbf{y},\bm{\varphi},\bm{\phi}}\min~f_1(\mathbf{x},\bm{\phi}) + \mathbb{I}(\mathbf{x},\bm{\alpha},\mathbf{y},\bm{\phi},\bm{\varphi})\\
&\quad~\text{s.t.}\quad~\mathbf{y} = \mathbf{x},\\
&\hspace{1.3 cm}\bm{\varphi} = \bm{\phi},
\end{align}
\end{subequations}
whose optimal solution can be obtained by minimizing its augmented Lagrangian (AL) function:
\begin{equation}\label{eq:AL function}\begin{aligned}
&\mathcal{L}(\mathbf{x},\bm{\alpha} ,\mathbf{y},\bm{\phi},\bm{\varphi},\bm{\mu}_1,\bm{\mu}_2) \triangleq f_1(\mathbf{x},\bm{\phi}) +\mathbb{I}(\mathbf{x},\bm{\alpha}, \mathbf{y},\bm{\phi},\bm{\varphi})\\
&\quad\quad\quad\quad + \frac{\rho}{2}\big\|\mathbf{x}-\mathbf{y}+\frac{\bm{\mu}_1}{\rho}\big\|^2 + \frac{\rho}{2}\big\|\bm{\phi}-\bm{\varphi}+\frac{\bm{\mu}_2}{\rho}\big\|^2,
\end{aligned}\end{equation}
where $\rho>0$ is a penalty parameter, and $\bm{\mu}_1\in\mathbb{C}^{ML}$, $\bm{\mu}_2\in\mathbb{C}^{N}$ are dual variables.
We observe that minimizing the AL function (\ref{eq:AL function}) is more tractable than the original problem (\ref{eq:problem3}) after removing equality constraints (\ref{eq:problem3}e) and (\ref{eq:problem3}f).
However, the complicated non-convex function $f_1(\mathbf{x},\bm{\phi})$ greatly hinders a direct solution.
Therefore, we propose to utilize the MM method to convert the AL minimization problem into a sequence of simpler problems to be solved until convergence.

\subsection{MM-based Transformation}

Specifically, given the obtained solutions $\mathbf{x}_t$ and $\bm{\phi}_t$ in the $t$-th iteration, we attempt to construct a more tractable surrogate function that approximates the objective function $\mathcal{L}(\mathbf{x},\bm{\alpha} ,\mathbf{y},\bm{\phi},\bm{\varphi},\bm{\mu}_1,\bm{\mu}_2)$ at the local point $(\mathbf{x}_t, \bm{\phi}_t)$ and serves as an upper-bound to be minimized in the next iteration.
As shown in the equation (14) in \cite{Sun TSP 17}, function $\mathbf{s}^H\mathbf{M}^{-1}\mathbf{s}$ is jointly convex with respect to $\mathbf{s}$ and positive-definite matrix $\mathbf{M}$, and a surrogate function of $\mathbf{s}^H\mathbf{M}^{-1}\mathbf{s}$ at point $(\mathbf{s}_t,\mathbf{M}_t)$ is given by
\begin{equation}\label{eq:surrogate}\non
\mathbf{s}^H\mathbf{M}^{-1}\mathbf{s} \geq 2\Re\big\{\mathbf{s}_t^H\mathbf{M}_t^{-1}\mathbf{s}\big\}-\text{Tr}\big\{\mathbf{M}_t^{-1}
\mathbf{s}_t\mathbf{s}_t^H\mathbf{M}_t^{-1}\mathbf{M}\big\}+c,
\end{equation}
where $c$ is a scalar that is irrelevant to the variables $\mathbf{s}$ and $\mathbf{M}$.
Therefore, by defining the affine relationship
\begin{subequations}\label{eq:st Mt}
\begin{align}
\mathbf{s} &\triangleq \widetilde{\mathbf{H}}_0(\bm{\phi})\mathbf{x},\\
\mathbf{M} &\triangleq \sum_{q=1}^Q\varsigma_q^2\widetilde{\mathbf{H}}_q(\bm{\phi})\mathbf{xx}^H
\widetilde{\mathbf{H}}^H_q(\bm{\phi})+\varsigma_\text{z}^2\mathbf{I},
\end{align}
\end{subequations}
a surrogate function of $f_1(\mathbf{x},\bm{\phi})$ can be expressed as
\begin{equation}\label{eq:f surrogate}
\begin{aligned}
&f_1(\mathbf{x},\bm{\phi})\leq -2\Re\big\{\mathbf{s}_t^H\mathbf{M}_t^{-1}\widetilde{\mathbf{H}}_0(\bm{\phi})\mathbf{x}\big\}+c_1\\
&~+\text{Tr}\Big\{\mathbf{M}_t^{-1}
\mathbf{s}_t\mathbf{s}_t^H\mathbf{M}_t^{-1}\Big[\sum_{q=1}^Q\varsigma_q^2\widetilde{\mathbf{H}}_q(\bm{\phi})\mathbf{xx}^H
\widetilde{\mathbf{H}}^H_q(\bm{\phi})\Big]\Big\},
\end{aligned}
\end{equation}
where $c_1$ is a constant irrelevant to the variables $\mathbf{x}$ and $\bm{\phi}$, $\mathbf{s}_t \triangleq \widetilde{\mathbf{H}}_0(\bm{\phi}_t)\mathbf{x}_t$, and $\mathbf{M}_t \triangleq \sum_{q=1}^Q\varsigma_q^2\widetilde{\mathbf{H}}_q(\bm{\phi}_t)\mathbf{x}_t\mathbf{x}_t^H
\widetilde{\mathbf{H}}^H_q(\bm{\phi}_t)+\varsigma_\text{z}^2\mathbf{I}$.
Plugging (\ref{eq:f surrogate}) into (\ref{eq:AL function}), a surrogate function for $\mathcal{L}(\mathbf{x},\bm{\alpha},\mathbf{y},\bm{\phi},\bm{\varphi},\bm{\mu}_1,\bm{\mu}_2)$ can be expressed as (\ref{eq:surrogate AL function}) presented at the top of the next page, which can be minimized by alternately updating $\mathbf{x}$ and $\bm{\alpha}$, $\mathbf{y}$, $\bm{\phi}$, $\bm{\varphi}$, $\bm{\mu}_1$, and $\bm{\mu}_2$ as shown in the following subsections.

\begin{figure*}
\begin{equation}\label{eq:surrogate AL function}\begin{aligned}
\mathcal{L}(\mathbf{x},\bm{\alpha},\mathbf{y},\bm{\phi},\bm{\varphi},\bm{\mu}_1,\bm{\mu}_2)\leq~& \text{Tr}\Big\{\mathbf{M}_t^{-1}
\mathbf{s}_t\mathbf{s}_t^H\mathbf{M}_t^{-1}\Big[\sum_{q=1}^Q\varsigma_q^2
\widetilde{\mathbf{H}}_q(\bm{\phi})\mathbf{xx}^H
\widetilde{\mathbf{H}}^H_q(\bm{\phi})\Big]\Big\} -2\Re\big\{\mathbf{s}_t^H\mathbf{M}_t^{-1}\widetilde{\mathbf{H}}_0(\bm{\phi})\mathbf{x}\big\} \\
& +\frac{\rho}{2}\big\|\mathbf{x}-\mathbf{y}+\frac{\bm{\mu}_1}{\rho}\big\|^2 + \frac{\rho}{2}\big\|\bm{\phi}-\bm{\varphi}+\frac{\bm{\mu}_2}{\rho}\big\|^2
+\mathbb{I}(\mathbf{x},\bm{\alpha},\mathbf{y},\bm{\phi},\bm{\varphi}) + c_1.
\end{aligned}\end{equation}
\rule[-0pt]{18.5 cm}{0.05em}
\end{figure*}

\subsection{Update $\mathbf{x}$ and $\bm{\alpha}$}

Given $\mathbf{y}_t$, $\bm{\phi}_t$, $\bm{\varphi}_t$, $\bm{\mu}_1$, and $\bm{\mu}_2$, the optimization problem for updating the transmit waveform $\mathbf{x}$ and the associated scaling factor $\bm{\alpha}$ to minimize the surrogate AL function (\ref{eq:surrogate AL function}) can be written as
\begin{equation}\label{eq:update x original}
\begin{aligned}
&\underset{\mathbf{x},\bm{\alpha}}{\min}~~\text{Tr}\Big\{\mathbf{M}_t^{-1}\mathbf{s}_t\mathbf{s}_t^H\mathbf{M}_t^{-1}
\Big[\sum_{q=1}^Q\varsigma_q^2
\widetilde{\mathbf{H}}_q(\bm{\phi}_t)\mathbf{xx}^H
\widetilde{\mathbf{H}}^H_q(\bm{\phi}_t)\Big]\Big\}\\
&\quad\quad~-2\Re\big\{\mathbf{s}_t^H\mathbf{M}_t^{-1}\widetilde{\mathbf{H}}_0(\bm{\phi}_t)\mathbf{x}\big\} + \frac{\rho}{2}\big\|\mathbf{x}-\mathbf{y}_t+\frac{\bm{\mu}_1}{\rho}\big\|^2\\
&\quad\quad~+\mathbb{I}(\mathbf{x},\bm{\alpha},\mathbf{y}_t,\bm{\phi}_t,\bm{\varphi}_t).
\end{aligned}
\end{equation}
For conciseness, we define
\begin{subequations}\label{eq:Dtdthkl}
\begin{align}
\mathbf{D}_t &\triangleq \hspace{-0.07 cm}\sum_{q=1}^Q\hspace{-0.07 cm}\varsigma_q^2\widetilde{\mathbf{H}}^H_q
\hspace{-0.05 cm}(\hspace{-0.03 cm}\bm{\phi}_t\hspace{-0.03 cm})\mathbf{M}_t^{-1}
\mathbf{s}_t\mathbf{s}_t^H\mathbf{M}_t^{-1}\widetilde{\mathbf{H}}_q\hspace{-0.05 cm}(\hspace{-0.03 cm}\bm{\phi}_t\hspace{-0.03 cm})
\hspace{-0.05 cm}+\hspace{-0.05 cm}\frac{\rho}{2}\mathbf{I}_{ML},\\
\mathbf{d}_t &\triangleq -2\widetilde{\mathbf{H}}^H_0(\bm{\phi}_t)\mathbf{M}_t^{-1}\mathbf{s}_t-\rho\mathbf{y}_t+\bm{\mu}_1,\\
\widetilde{\mathbf{h}}_{k,l} &\triangleq \mathbf{h}^H_{k,l}+(\mathbf{e}_l^T\otimes\bm{\phi}_t^T)\mathbf{G}_{k}.
\end{align}
\end{subequations}
Then, by dropping the constant terms that are irrelevant to the variables $\mathbf{x}$ and $\bm{\alpha}$, problem (\ref{eq:update x original}) can be re-formulated as
\begin{subequations}\label{eq:update x}
\begin{align}
&\underset{\mathbf{x},\bm{\alpha}}\min~~~\mathbf{x}^H\mathbf{D}_t\mathbf{x}+\Re\big\{\mathbf{d}_t^H\mathbf{x}\big\}\\
&~\text{s.t.}~~~\widetilde{\mathbf{h}}_{k,l}^H\mathbf{x} = \alpha_{k,l}\gamma_{k,l}^\text{ZF},~~\alpha_{k,l} \geq 1,~~\forall k,l, \\
&~~~~~~~\text{or}~~\big|\widetilde{\mathbf{h}}_{k,l}^H\mathbf{x}- \alpha_{k,l}\gamma_{k,l}^\text{ZF}\big|^2 \leq \epsilon,~~\alpha_{k,l} \geq 1, ~~\forall k,l,\\
&~~~~~~~\text{or}~~\Re\big\{\gamma_{k,l}^\text{CI}\widetilde{\mathbf{h}}_{k,l}^H\mathbf{x}\big\}
\geq 1,~~\forall k,l, \\
&~~~~~~~|x_i| \leq \sqrt{P/M},~~\forall i.
\end{align}
\end{subequations}
It is obvious that problem (\ref{eq:update x}) is convex for all three communication QoS constraints (\ref{eq:update x}b)-(\ref{eq:update x}d), and thus can be optimally solved by various existing algorithms or toolbox, e.g., CVX.

\subsection{Update $\mathbf{y}$}

With obtained $\mathbf{x}_t$, $\bm{\alpha}_t$, $\bm{\phi}_t$, $\bm{\varphi}_t$, $\bm{\mu}_1$, and $\bm{\mu}_2$, the optimization problem of solving for the auxiliary variable $\mathbf{y}$ can be expressed as
\begin{equation}
\underset{\mathbf{y}}\min~~\frac{\rho}{2}\big\|\mathbf{x}_t-\mathbf{y}+\frac{\bm{\mu}_1}{\rho}\big\|^2
+ \mathbb{I}(\mathbf{x}_t,\bm{\alpha}_t,\mathbf{y},\bm{\phi}_t,\bm{\varphi}_t),
\end{equation}
which is equivalent to
\begin{subequations}
\begin{align}
&\underset{\mathbf{y}}\min~~\big\|\mathbf{x}_t-\mathbf{y}+\frac{\bm{\mu}_1}{\rho}\big\|^2\\
&~\text{s.t.}~~~|y_i| = \sqrt{P/M},~~\forall i.
\end{align}
\end{subequations}
Thus, the optimal solution of $\mathbf{y}$ can be easily obtained by the phase alignment:
\begin{equation}\label{eq:ystar}
\mathbf{y}^\star = \sqrt{P/M}e^{\jmath\angle(\rho\mathbf{x}_t+\bm{\mu}_1)}.
\end{equation}

\subsection{Update $\bm{\phi}$}

After obtaining $\mathbf{x}_t$, $\bm{\alpha}_t$, $\mathbf{y}_t$, $\bm{\varphi}_t$, $\bm{\mu}_1$, and $\bm{\mu}_2$, the problem of optimizing the reflecting coefficients $\bm{\phi}$ can be formulated as
\begin{equation}\label{eq:problem phi original}
\begin{aligned}
&\underset{\bm{\phi}}\min~~\text{Tr}\Big\{\mathbf{M}_t^{-1}
\mathbf{s}_t\mathbf{s}_t^H\mathbf{M}_t^{-1}\Big[\sum_{q=1}^Q\varsigma_q^2
\widetilde{\mathbf{H}}_q(\bm{\phi})\mathbf{x}_t\mathbf{x}_t^H
\widetilde{\mathbf{H}}^H_q(\bm{\phi})\Big]\Big\}\\
&\quad\quad~-2\Re\big\{\mathbf{s}_t^H\mathbf{M}_t^{-1}\widetilde{\mathbf{H}}_0(\bm{\phi})\mathbf{x}_t\big\}
 + \frac{\rho}{2}\big\|\bm{\phi}-\bm{\varphi}_t+\frac{\bm{\mu}_2}{\rho}\big\|^2 \\
&\quad\quad~ + \mathbb{I}(\mathbf{x}_t,\bm{\alpha}_t,\mathbf{y}_t,\bm{\phi},\bm{\varphi}_t).
\end{aligned}
\end{equation}

\subsubsection{Reformulation}

To facilitate the algorithm development, we first utilize the definition of $\mathbb{I}(\mathbf{x}_t,\bm{\alpha}_t,\mathbf{y}_t,\bm{\phi},\bm{\varphi}_t)$ to equivalently transform (\ref{eq:problem phi original}) into
\begin{subequations}\label{eq:problem phi}
\begin{align}
&\underset{\bm{\phi}}\min~~f_2(\bm{\phi})\\
&~\text{s.t.}~~(\ref{eq:communication QoS constraints}\text{a})~\text{or}~(\ref{eq:communication QoS constraints}\text{b})~\text{or}~ (\ref{eq:communication QoS constraints}\text{c}),\\
&\qquad~|\phi_n| \leq 1,~\forall n,
\end{align}
\end{subequations}
where for brevity we define
\begin{equation}\label{eq:fphi}\begin{aligned}
&f_2(\bm{\phi}) \triangleq \text{Tr}\Big\{\mathbf{M}_t^{-1}
\mathbf{s}_t\mathbf{s}_t^H\mathbf{M}_t^{-1}\Big[\sum_{q=1}^Q\varsigma_q^2
\widetilde{\mathbf{H}}_q(\bm{\phi})\mathbf{x}_t\mathbf{x}_t^H
\widetilde{\mathbf{H}}^H_q(\bm{\phi})\Big]\Big\}\\
&\quad\quad\quad\quad-2\Re\big\{\mathbf{s}_t^H\mathbf{M}_t^{-1}\widetilde{\mathbf{H}}_0(\bm{\phi})\mathbf{x}_t\big\}
 +  \frac{\rho}{2}\big\|\bm{\phi}-\bm{\varphi}_t+\frac{\bm{\mu}_2}{\rho}\big\|^2.
\end{aligned}\end{equation}
\nid In problem (\ref{eq:problem phi}), we observe that the variable $\bm{\phi}$ is embedded in the functions $\widetilde{\mathbf{H}}_0(\bm{\phi})$, $\widetilde{\mathbf{H}}_q(\bm{\phi}), \forall q$, and some Kronecker product terms, which are not amenable for optimization.
Therefore, in order to facilitate the algorithm development, we first equivalently recast problem (\ref{eq:problem phi}) into an explicit problem as
\begin{subequations}\label{eq:problem phi reform}
\begin{align}
&\underset{\bm{\phi}}\min~f_2(\bm{\phi}) = \bm{\phi}^H\mathbf{F}_t\bm{\phi}+\Re\{\bm{\phi}^H\mathbf{f}_t\}
+\mathbf{v}^H\mathbf{F}_{\text{v},t}\mathbf{v}+\Re\{\mathbf{v}^H\mathbf{f}_{\text{v},t}\} \non \\
&\hspace{1.6 cm} + \Re\{\mathbf{v}^H\mathbf{L}_t\bm{\phi}\} + c_2 + \frac{\rho}{2}\big\|\bm{\phi}-\bm{\varphi}_t+\frac{\bm{\mu}_2}{\rho}\big\|^2\\
&\text{s.t.}~\mathbf{h}^H_{k,l}\mathbf{x} + \widetilde{\mathbf{g}}_{k,l}^T\bm{\phi} =\alpha_{k,l}\gamma_{k,l}^\text{ZF},~~\alpha_{k,l} \geq 1,~~\forall k,l, \\
&\quad~\text{or}~~\big|\mathbf{h}^H_{k,l}\mathbf{x}\hspace{-0.05 cm}+\hspace{-0.05 cm}\widetilde{\mathbf{g}}_{k,l}^T\bm{\phi} \hspace{-0.05 cm}-\hspace{-0.05 cm}\alpha_{k,l}\gamma_{k,l}^\text{ZF}\big|^2\hspace{-0.05 cm}\leq\hspace{-0.05 cm}\epsilon,~\alpha_{k,l}\hspace{-0.05 cm}\geq\hspace{-0.05 cm}1,~\forall k,l,\\
&\quad~\text{or}~~\Re\big\{\gamma_{k,l}^\text{CI}(\mathbf{h}^H_{k,l}\mathbf{x} + \widetilde{\mathbf{g}}_{k,l}^T\bm{\phi})\big\} \geq 1,~~\forall k,l,\\
&\quad~\left|\phi_n\right| \leq 1,~~\forall n,
\end{align}
\end{subequations}
where $\mathbf{v} \triangleq \text{vec}\{\bm{\phi\phi}^T\} = \bm{\phi}\otimes \bm{\phi}$.
The equivalence between problem (\ref{eq:problem phi}) and problem (\ref{eq:problem phi reform}) is proved in Appendix A, in which the expressions of $\mathbf{F}_t$, $\mathbf{f}_t$, $\mathbf{F}_{\text{v},t}$, $\mathbf{f}_{\text{v},t}$, and $\mathbf{L}_t$ are shown in (\ref{eq:fl reform}) and the matrices $\mathbf{F}_t$ and $\mathbf{F}_{\text{v},t}$ are positive semi-definite.
It can be observed that problem (\ref{eq:problem phi reform}) is non-convex due to the non-convex terms $\mathbf{v}^H\mathbf{F}_{\text{v},t}\mathbf{v}$, $\Re\{\mathbf{v}^H\mathbf{f}_{\text{v},t}\}$, and $\Re\{\mathbf{v}^H\mathbf{L}_t\bm{\phi}\}$ in the objective function.
In order to handle the non-convex objective function $f_2(\bm{\phi})$, in the following we propose to seek for a more tractable surrogate function by applying the second-order Taylor expansion after some matrix manipulations.

\subsubsection{MM-based Transformation}

Specifically, since $\mathbf{F}_{\text{v},t}$ is a positive semidefinite Hermitian matrix according to its definition in (\ref{eq:fl reform}c), a surrogate function of $\mathbf{v}^H\mathbf{F}_{\text{v},t}\mathbf{v}$ with respect to $\mathbf{v}$ can be obtained as
\begin{equation}\label{eq:sur vFv}\begin{aligned}
\mathbf{v}^H\mathbf{F}_{\text{v},t}\mathbf{v} &\leq\lambda_1\mathbf{v}^H\mathbf{v}
+ 2\Re\big\{\mathbf{v}^H(\mathbf{F}_{\text{v},t}-\lambda_1\mathbf{I}_{N^2})\mathbf{v}_t\big\} \\
&\quad + \mathbf{v}_t^H(\lambda_1\mathbf{I}_{N^2}-\mathbf{F}_{\text{v},t})\mathbf{v}_t,
\end{aligned}\end{equation}
where $\lambda_1$ is an upper-bound of the eigenvalues of matrix $\mathbf{F}_{\text{v},t}$.
Considering the prohibitively high complexity caused by the eigen decomposition of an $N^2\times N^2$ matrix, we choose $\lambda_1= \text{Tr}\{\mathbf{F}_{\text{v},t}\}$ as an efficient solution, which is essentially the summation of all eigenvalues.
Since the matrix $\mathbf{F}_{\text{v},t}$ contains $Q$ rank-one matrices as expressed in (\ref{eq:fl reform}c), instead of computing and storing this $N^2\times N^2$-dimensional matrix $\mathbf{F}_{\text{v},t}$, we can directly obtain its trace by
\be
\lambda_1 = \text{Tr}\{\mathbf{F}_{\text{v},t}\} = 2\sum_{q=1}^Q\varsigma_q^2\|\widetilde{\mathbf{D}}_{q,t}\|_F^2,
\ee
where the expression of $\widetilde{\mathbf{D}}_{q,t}\in\mathbb{C}^{N\times N}$ is given in (\ref{eq:Dqcq}a).
In addition, thanks to the amplitude constraint (\ref{eq:problem phi reform}e), the quadratic term $\mathbf{v}^H\mathbf{v}$ is upper-bounded by
\be
\mathbf{v}^H\mathbf{v} = (\bm{\phi}\otimes\bm{\phi})^H(\bm{\phi}\otimes\bm{\phi})
= (\bm{\phi}^H\bm{\phi})\otimes(\bm{\phi}^H\bm{\phi}) \leq N^2.
\ee
Thus, a surrogate function of $\mathbf{v}^H\mathbf{F}_{\text{v},t}\mathbf{v}$ with respect to $\mathbf{v}$ is given by
\begin{equation}\label{eq:vFv}
\mathbf{v}^H\mathbf{F}_{\text{v},t}\mathbf{v} \leq\Re\{\mathbf{v}^H\widetilde{\mathbf{f}}_{\text{v},t}\} + c_3,
\end{equation}
where we define
\be\label{eq:fvttilde}
\widetilde{\mathbf{f}}_{\text{v},t} \triangleq 2(\mathbf{F}_{\text{v},t}-\lambda_1\mathbf{I}_{N^2})\mathbf{v}_t,
\ee
and the scalar $c_3$ is irrelevant to the variable $\mathbf{v}$ (i.e., $\bm{\phi}$).

Then, adding the forth term in (\ref{eq:problem phi reform}a) to (\ref{eq:vFv}) and utilizing $\mathbf{v}\triangleq\bm{\phi}\otimes\bm{\phi}$, we have
\begin{subequations}\label{eq:vFv+vf}
\begin{align}
\mathbf{v}^H\mathbf{F}_{\text{v},t}\mathbf{v}+\Re\{\mathbf{v}^H\mathbf{f}_{\text{v},t}\} &\leq
\Re\{\mathbf{v}^H(\widetilde{\mathbf{f}}_{\text{v},t}+\mathbf{f}_{\text{v},t})\} + c_3 \\
& = \Re\{\bm{\phi}^H\widetilde{\mathbf{F}}_{\text{v},t}\bm{\phi}^*\} + c_3,
\end{align}
\end{subequations}
where $\widetilde{\mathbf{F}}_{\text{v},t}\in\mathbb{C}^{N\times N}$ is a reshaped version of $\widetilde{\mathbf{f}}_{\text{v},t}+\mathbf{f}_{\text{v},t}$, i.e., $\widetilde{\mathbf{f}}_{\text{v},t}+\mathbf{f}_{\text{v},t} = \text{vec}\{\widetilde{\mathbf{F}}_{\text{v},t}\}$.
In an effort to provide an efficient solution, we attempt to avoid the computation of $\widetilde{\mathbf{f}}_{\text{v},t}$ which involves the $N^2\times N^2$-dimensional matrix $\mathbf{F}_{\text{v},t}$ as expressed in (\ref{eq:fvttilde}), and propose to directly construct $\widetilde{\mathbf{F}}_{\text{v},t}$ using $N\times N$ lower-dimensional matrices as
\be
\widetilde{\mathbf{F}}_{\text{v},t}\hspace{-0.06cm}= \hspace{-0.06cm} 4\hspace{-0.08cm}\sum_{q=1}^Q\hspace{-0.08cm}\varsigma_q^2(\bm{\phi}_t^T\widetilde{\mathbf{D}}_{q,t}^H\bm{\phi}_t\hspace{-0.03cm}
+\hspace{-0.03cm}\mathbf{s}_t^H\mathbf{M}_t^{-1}\mathbf{a}_q)\widetilde{\mathbf{D}}_{q,t}
\hspace{-0.03cm}-\hspace{-0.03cm}2\widetilde{\mathbf{D}}_{0,t}\hspace{-0.03cm}-\hspace{-0.03cm}2\lambda_1\bm{\phi}_t\bm{\phi}_t^T,
\ee
whose equivalence is proved in Appendix B.

Since the real-valued function $\Re\{\bm{\phi}^H\widetilde{\mathbf{F}}_{\text{v},t}\bm{\phi}^*\}$ in (\ref{eq:vFv+vf}b) is still non-convex with respect to $\bm{\phi}$, we further propose to convert the complex-valued variable into its real-valued form and utilize the second-order Taylor expansion to seek for a convex surrogate function.
Specifically, by defining
\begin{subequations}
\begin{align}\label{eq:phibar}
\overline{\bm{\phi}} &\triangleq [\Re\{\bm{\phi}^T\}~~\Im\{\bm{\phi}^T\}]^T,\\
\overline{\mathbf{F}}_{\text{v},t} &\triangleq \left[\begin{array}{cc}\Re\{\widetilde{\mathbf{F}}_{\text{v},t}\} &\Im\{\widetilde{\mathbf{F}}_{\text{v},t}\} \\ \Im\{\widetilde{\mathbf{F}}_{\text{v},t}\} & -\Re\{\widetilde{\mathbf{F}}_{\text{v},t}\}\end{array}\right],
\end{align}
\end{subequations}
a convex surrogate function can be derived as
\begin{subequations}\label{eq:phiFphi}
\begin{align}
\Re\{\bm{\phi}^H\widetilde{\mathbf{F}}_{\text{v},t}\bm{\phi}^*\} & = \overline{\bm{\phi}}^T\overline{\mathbf{F}}_{\text{v},t}\overline{\bm{\phi}}\\
& \leq \overline{\bm{\phi}}_t^T\overline{\mathbf{F}}_{\text{v},t}\overline{\bm{\phi}}_t + \overline{\bm{\phi}}_t^T(\overline{\mathbf{F}}_{\text{v},t}+\overline{\mathbf{F}}_{\text{v},t}^T)
(\overline{\bm{\phi}}-\overline{\bm{\phi}}_t)\non \\
&\quad + \frac{\lambda_2}{2}(\overline{\bm{\phi}}-\overline{\bm{\phi}}_t)^T
(\overline{\bm{\phi}}-\overline{\bm{\phi}}_t)\\
& = \frac{\lambda_2}{2}\overline{\bm{\phi}}^T\overline{\bm{\phi}} + \Re\{\overline{\bm{\phi}}^T\overline{\mathbf{f}}_{\text{v},t}\} + c_4\\
& = \frac{\lambda_2}{2}\bm{\phi}^H\bm{\phi} + \Re\{\bm{\phi}^H\mathbf{U}\overline{\mathbf{f}}_{\text{v},t}\} + c_4,
\end{align}
\end{subequations}
where $\lambda_2$ is the maximum eigenvalue of the Hessian matrix $(\overline{\mathbf{F}}_{\text{v},t}+\overline{\mathbf{F}}_{\text{v},t}^T)$, $\overline{\mathbf{f}}_{\text{v},t}\triangleq (\overline{\mathbf{F}}_{\text{v},t}+\overline{\mathbf{F}}_{\text{v},t}^T-\lambda_2\mathbf{I}_{2N})\overline{\bm{\phi}}_t$, the scalar $c_4$ is irrelevant to the variable $\overline{\bm{\phi}}$ (i.e., $\bm{\phi}$), and we define $\mathbf{U} \triangleq [\mathbf{I}_N~\jmath\mathbf{I}_N]$ to realize $\Re\{\overline{\bm{\phi}}^T\overline{\mathbf{f}}_{\text{v},t}\} = \Re\{\bm{\phi}^H\mathbf{U}\overline{\mathbf{f}}_{\text{v},t}\}$.
Based on the derivations in (\ref{eq:vFv+vf}) and (\ref{eq:phiFphi}), we have
\begin{equation}\label{eq:sur1}
\mathbf{v}^H\mathbf{F}_{\text{v},t}\mathbf{v}\!+\!\Re\{\mathbf{v}^H\mathbf{f}_{\text{v},t}\} \leq \frac{\lambda_2}{2}\bm{\phi}^H\bm{\phi}+ \! \Re\{\bm{\phi}^H\mathbf{U}\overline{\mathbf{f}}_{\text{v},t}\} + \!c_3 + \!c_4,\!
\end{equation}
which provides a convex surrogate function to the third-plus-forth term in the objective function (\ref{eq:problem phi reform}a).

Then, we turn to find a tractable surrogate function for the fifth term $\Re\{\mathbf{v}^H\mathbf{L}_t\bm{\phi}\}$ in (\ref{eq:problem phi reform}a).
We first apply the expression of $\mathbf{L}_t$ in (\ref{eq:fl reform}e) and $\mathbf{v} = \bm{\phi}\otimes\bm{\phi}$ to explicitly re-write the term $\Re\{\mathbf{v}^H\mathbf{L}_t\bm{\phi}\}$ with respect to $\bm{\phi}$ as
\begin{subequations}
\begin{align}
\Re\big\{\mathbf{v}^H\mathbf{L}_t\bm{\phi}\big\} &\hspace{-0.03 cm}= \hspace{-0.03 cm}4\hspace{-0.06 cm}\sum_{q=1}^Q\hspace{-0.06 cm}\varsigma_q^2\Re\big\{\hspace{-0.06 cm}(\bm{\phi}^H\hspace{-0.06 cm}\otimes\hspace{-0.06 cm}\bm{\phi}^H)\text{vec}\{\widetilde{\mathbf{D}}_{q,t}\}\widetilde{\mathbf{c}}_{q,t}^H\bm{\phi}\big\}\\
& \hspace{-0.03 cm}=\hspace{-0.03 cm}4\sum_{q=1}^Q\varsigma_q^2\Re\big\{\widetilde{\mathbf{c}}_{q,t}^H\bm{\phi}\bm{\phi}^H\widetilde{\mathbf{D}}_{q,t}\bm{\phi}^*\big\}.
\end{align}
\end{subequations}
Using the definition (\ref{eq:phibar}) and defining
\begin{subequations}
\begin{align}
\overline{\mathbf{c}}_{q,t,1} &\triangleq [\Re\{\widetilde{\mathbf{c}}_{q,t}^T\}~~\Im\{\widetilde{\mathbf{c}}_{q,t}^T\}]^T,\\
\overline{\mathbf{c}}_{q,t,2} &\triangleq [\Im\{\widetilde{\mathbf{c}}_{q,t}^T\}~~-\Re\{\widetilde{\mathbf{c}}_{q,t}^T\}]^T,\\
\overline{\mathbf{D}}_{q,t,1} &\triangleq \left[\begin{array}{cc}\Re\{\widetilde{\mathbf{D}}_{q,t}\} &\Im\{\widetilde{\mathbf{D}}_{q,t}\} \\ \Im\{\widetilde{\mathbf{D}}_{q,t}\} & -\Re\{\widetilde{\mathbf{D}}_{q,t}\}\end{array}\right],\\
\overline{\mathbf{D}}_{q,t,2} &\triangleq \left[\begin{array}{cc}\Im\{\widetilde{\mathbf{D}}_{q,t}\} &-\Re\{\widetilde{\mathbf{D}}_{q,t}\} \\ -\Re\{\widetilde{\mathbf{D}}_{q,t}\} & -\Im\{\widetilde{\mathbf{D}}_{q,t}\}\end{array}\right],
\end{align}
\end{subequations}
the real-valued form of $\Re\{\mathbf{v}^H\mathbf{L}_t\bm{\phi}\}$ can be expressed as
\begin{equation}\label{eq:phiLv}\begin{aligned}
f_3(\overline{\bm{\phi}})&\triangleq\Re\{\mathbf{v}^H\mathbf{L}_t\bm{\phi}\} \\
&=4\sum_{q=1}^Q\varsigma_q^2\sum_{i=1}^2\overline{\mathbf{c}}_{q,t,i}^T\overline{\bm{\phi}}\overline{\bm{\phi}}^T
\overline{\mathbf{D}}_{q,t,i}\overline{\bm{\phi}},
\end{aligned}\end{equation}
whose first-order and second-order derivatives can be calculated as
\begin{subequations}\label{eq:derivatives}
\begin{align}
\nabla f_3(\overline{\bm{\phi}}) &= 4\sum_{q=1}^Q\varsigma_q^2\sum_{i=1}^2\big[\overline{\mathbf{c}}_{q,t,i}^T\overline{\bm{\phi}}
(\overline{\mathbf{D}}_{q,t,i}+\overline{\mathbf{D}}_{q,t,i}^T)\overline{\bm{\phi}} \non\\
&\quad+ \overline{\bm{\phi}}^T
\overline{\mathbf{D}}_{q,t,i}\overline{\bm{\phi}}\overline{\mathbf{c}}_{q,t,i}\big],\\
\nabla^2 f_3(\overline{\bm{\phi}}) &=4\sum_{q=1}^Q\varsigma_q^2\sum_{i=1}^2\big[(\overline{\mathbf{D}}_{q,t,i}+\overline{\mathbf{D}}_{q,t,i}^T)\overline{\bm{\phi}}\overline{\mathbf{c}}_{q,t,i}^T\non\\
&\quad+(\overline{\mathbf{c}}_{q,t,i}\overline{\bm{\phi}}^T
\hspace{-0.05 cm}+\hspace{-0.05 cm}\overline{\bm{\phi}}^T\overline{\mathbf{c}}_{q,t,i}\mathbf{I})(\overline{\mathbf{D}}_{q,t,i}\hspace{-0.05 cm}+\hspace{-0.05 cm}\overline{\mathbf{D}}_{q,t,i}^T)
\big].
\end{align}
\end{subequations}
With $\nabla f_3(\overline{\bm{\phi}})$ and $\nabla^2 f_3(\overline{\bm{\phi}})$ shown in (\ref{eq:derivatives}), a surrogate function of $f_3(\overline{\bm{\phi}})$ can be obtained by
\begin{subequations}\label{eq:f3s}
\begin{align}
f_3(\overline{\bm{\phi}}) &\leq f_3(\overline{\bm{\phi}}_t) + (\nabla f_3(\overline{\bm{\phi}}_t))^T
(\overline{\bm{\phi}}-\overline{\bm{\phi}}_t) \non \\
&\quad~ + \frac{\lambda_3}{2}(\overline{\bm{\phi}}-\overline{\bm{\phi}}_t)^T
(\overline{\bm{\phi}}-\overline{\bm{\phi}}_t)\\
&= \frac{\lambda_3}{2}\overline{\bm{\phi}}^T\overline{\bm{\phi}} + \overline{\bm{\phi}}^T\overline{\bm{\ell}}_t + c_5,\\
& = \frac{\lambda_3}{2}\bm{\phi}^H\bm{\phi} + \Re\{\bm{\phi}^H\mathbf{U}\overline{\bm{\ell}}_t\} + c_5,
\end{align}
\end{subequations}
where $\lambda_3$ is the maximum eigenvalue of the Hessian matrix $\nabla^2 f_3(\overline{\bm{\phi}})$, $\overline{\bm{\ell}}_t \triangleq \nabla f_3(\overline{\bm{\phi}}_t)-\lambda_3\overline{\bm{\phi}}_t$, and the scalar $c_5$ is irrelevant to the variable $\overline{\bm{\phi}}$ (i.e., $\bm{\phi}$).

Given the surrogate functions derived for $\mathbf{v}^H\mathbf{F}_{\text{v},t}\mathbf{v} + \Re\{\mathbf{v}^H\mathbf{f}_{\text{v},t}\}$ and $\Re\{\bm{\phi}^H\mathbf{L}_t\mathbf{v}\}$ in (\ref{eq:sur1}) and (\ref{eq:f3s}c), respectively,  a surrogate objective function of (\ref{eq:problem phi reform}a) can be expressed as
\begin{subequations}
\begin{align}
&f_2(\bm{\phi})\leq \bm{\phi}^H\mathbf{F}_t\bm{\phi}\hspace{-0.05 cm}+\hspace{-0.05 cm}\Re\{\bm{\phi}^H\mathbf{f}_t\}\hspace{-0.05 cm}+\hspace{-0.05 cm}\frac{\lambda_2 }{2}\bm{\phi}^H\bm{\phi}\hspace{-0.05 cm}+\hspace{-0.05 cm}\Re\{\bm{\phi}^H\mathbf{U}\overline{\mathbf{f}}_{\text{v},t}\}\hspace{-0.05 cm}+\hspace{-0.05 cm}c_3\non \\
&\hspace{-0.05 cm}+\hspace{-0.05 cm}c_4\hspace{-0.05 cm}+\hspace{-0.05 cm}\frac{\lambda_3}{2}\bm{\phi}^H\bm{\phi} \hspace{-0.05 cm}+\hspace{-0.05 cm} \Re\{\bm{\phi}^H\mathbf{U}\overline{\bm{\ell}}_t\} \hspace{-0.05 cm}+\hspace{-0.05 cm} c_5 \hspace{-0.05 cm}+\hspace{-0.05 cm} c_2 \hspace{-0.05 cm}+\hspace{-0.05 cm}\frac{\rho}{2}\big\|\bm{\phi}-\bm{\varphi}_t+\frac{\bm{\mu}_2}{\rho}\big\|^2\\
& = \bm{\phi}^H\widetilde{\mathbf{F}}_t\bm{\phi} + \Re\{\bm{\phi}^H\widetilde{\mathbf{f}}_t\} + c_6,
\end{align}
\end{subequations}
where for simplicity we define $\widetilde{\mathbf{F}}_t\triangleq \mathbf{F}_t + \frac{\lambda_2 + \lambda_3+\rho}{2}\mathbf{I}_N$, $\widetilde{\mathbf{f}}_t\triangleq \mathbf{f}_t+\mathbf{U}\overline{\mathbf{f}}_{\text{v},t} + \mathbf{U}\overline{\bm{\ell}}_t -\rho\bm{\varphi}_t+\bm{\mu}_2$, and $c_6\triangleq c_2 + c_3 + c_4 + c_5 + \frac{\rho}{2}\|\bm{\varphi}_t-\bm{\mu}_2/\rho\|^2$.
Therefore, the optimization problem for updating $\bm{\phi}$ is formulated as
\begin{subequations}\label{eq:update phi}
\begin{align}
&\underset{\bm{\phi}}\min~~\bm{\phi}^H\widetilde{\mathbf{F}}_t\bm{\phi} + \Re\{\bm{\phi}^H\widetilde{\mathbf{f}}_t\}\\
&\text{s.t.}~\mathbf{h}^H_{k,l}\mathbf{x} + \widetilde{\mathbf{g}}_{k,l}^T\bm{\phi} =\alpha_{k,l}\gamma_{k,l}^\text{ZF},~~\alpha_{k,l} \geq 1,~~\forall k,l, \\
&\quad~\text{or}~~\big|\mathbf{h}^H_{k,l}\mathbf{x}\hspace{-0.05 cm}+\hspace{-0.05 cm}\widetilde{\mathbf{g}}_{k,l}^T\bm{\phi} \hspace{-0.05 cm}-\hspace{-0.05 cm}\alpha_{k,l}\gamma_{k,l}^\text{ZF}\big|^2\hspace{-0.05 cm}\leq\hspace{-0.05 cm}\epsilon,~\alpha_{k,l}\hspace{-0.05 cm}\geq\hspace{-0.05 cm}1,~\forall k,l,\\
&\quad~\text{or}~~\Re\big\{\gamma_{k,l}^\text{CI}(\mathbf{h}^H_{k,l}\mathbf{x} + \widetilde{\mathbf{g}}_{k,l}^T\bm{\phi})\big\} \geq 1,~~\forall k,l,\\
&\quad~|\phi_n| \leq 1,~~\forall n,
\end{align}
\end{subequations}
which is a convex problem with any one of the communication constraints (\ref{eq:update phi}b)-(\ref{eq:update phi}d), and thus can be efficiently solved.

\subsection{Update $\bm{\varphi}$}

Fixing $\mathbf{x}_t$, $\bm{\alpha}_t$, $\mathbf{y}_t$, $\bm{\phi}_t$, $\bm{\mu}_1$, and $\bm{\mu}_2$, the auxiliary variable $\bm{\varphi}$ is updated in a similar way to $\mathbf{y}$ as
\begin{equation}\label{eq:varphi star}
\bm{\varphi}^\star = e^{\jmath\angle(\rho\bm{\phi}_t+\bm{\mu}_2)}.
\end{equation}

\subsection{Update $\bm{\mu}_1$ and $\bm{\mu}_2$}

After obtaining $\mathbf{x}_t$, $\bm{\alpha}_t$, $\mathbf{y}_t$, $\bm{\phi}_t$, and $\bm{\varphi}_t$, the dual variables $\bm{\mu}_1$ and $\bm{\mu}_2$ are updated by
\begin{subequations}\label{eq:update mu}
\begin{align}
\bm{\mu}_1 &:= \bm{\mu}_1 + \rho(\mathbf{x}_t-\mathbf{y}_t),\\
\bm{\mu}_2 &:= \bm{\mu}_2 + \rho(\bm{\phi}_t-\bm{\varphi}_t).
\end{align}
\end{subequations}

\subsection{Initialization}

For the above described ADMM-MM based alternating algorithm, a good starting point can greatly accelerate the convergence and boost the performance.
Therefore, we investigate to properly initialize the RIS reflecting coefficients $\bm{\phi}$ and the transmit waveform $\mathbf{x}$.
Generally speaking, the purpose of deploying RIS is to create more favorable radio environments by improving the quality of preferred channels as well as degrading the quality of undesired channels.
Thus, without considering the transmit waveform design at the BS, we utilize channel gain as the metric to initialize $\bm{\phi}$.
Specifically, we aim to maximize the channel gains of the target and the users minus the channel gains of the clutter sources, subject to the constant-modulus constraint of each reflecting element.
The optimization problem for initializing $\bm{\phi}$ is formulated as
\begin{subequations}\label{eq:initialize phi}
\begin{align}
&\underset{\bm{\phi}}\max~~\big\|\mathbf{h}^H_\text{t}+\mathbf{h}^H_\text{rt}\bm{\Phi}\mathbf{G}\big\|^2 + \sum_{k=1}^K\left\|\mathbf{h}_k^H+\mathbf{h}_{\text{r},k}^H\bm{\Phi}\mathbf{G}\right\|^2\non\\
&\quad\quad\quad-\sum_{q=1}^Q\left\|\mathbf{h}_{\text{c},q}^H+\mathbf{h}_{\text{rc},q}^H\bm{\Phi}\mathbf{G}\right\|^2
\\
&~\text{s.t.}~~~\left|\phi_n\right| = 1,~~\forall n,
\end{align}
\end{subequations}
where the objective is a non-convex smooth quadratic function with respect to $\bm{\phi}$ and the unit modulus constraint defines a Riemannian manifold.
This is a very typical problem in the field of RIS and can be efficiently solved by various algorithms, such as the manifold optimization in \cite{Yu ICCC 2019}.

After initializing $\bm{\phi}$, the effective channel $\widetilde{\mathbf{h}}_{k,l}$ is determined as in (\ref{eq:Dtdthkl}c).
In order to reserve sufficient flexibilities for maximizing radar output SINR under given communication QoS constraints, we propose to initialize the transmit waveform $\mathbf{x}$ by maximizing the minimum QoS of the users with available transmit power.
Therefore, the optimization problem for the ZF/MMSE-type communication constraints is formulated as
\begin{subequations}\label{eq:initialize x}
\begin{align}
&\underset{\mathbf{x}}\max~~\underset{k,l}\min~~\alpha_{k,l}
\\
&~\text{s.t.}~\big[\mathbf{h}_{k,l}^H \hspace{-0.05 cm}+\hspace{-0.05 cm}(\mathbf{e}_l^T\hspace{-0.05 cm}\otimes\hspace{-0.05 cm}\bm{\phi}^T)\mathbf{G}_{k}\big]\mathbf{x} = \alpha_{k,l}\gamma_{k,l}^\text{ZF},~\forall k,l,\\
&\quad\quad\text{or}~\big|[\mathbf{h}_{k,l}^H \hspace{-0.07 cm}+\hspace{-0.07 cm} (\mathbf{e}_l^T\hspace{-0.07 cm}\otimes\hspace{-0.07 cm}\bm{\phi}^T)\mathbf{G}_{k}]\mathbf{x}\hspace{-0.05 cm} -\hspace{-0.05 cm}\alpha_{k,l}\gamma_{k,l}^\text{ZF}\big|^2\hspace{-0.07 cm}\leq\hspace{-0.07 cm}\epsilon,~\forall k,l,\\
&~~~~~~|x_i| \leq \sqrt{P/M},~~\forall i,
\end{align}
\end{subequations}
where the power constraint (\ref{eq:initialize x}d) is a relaxed convex version of the constant-modulus constraint (\ref{eq:power constraint}) for the purpose of simplifying the optimization.
Similarly, the optimization problem for CI-type communication QoS constraint is expressed as
\begin{subequations}\label{eq:initialize x2}
\begin{align}
&\underset{\mathbf{x}}\max~~\underset{k,l}\min~~\Re\big\{\widetilde{\mathbf{h}}_{k,l}^H\mathbf{x}\big\}
\\
&~\text{s.t.}~~~|x_i| \leq \sqrt{P/M},~~\forall i.
\end{align}
\end{subequations}
It is obvious that the convex problems (\ref{eq:initialize x}) and (\ref{eq:initialize x2}) can be easily solved and the solutions after normalization can be used as the initialization for the alternating algorithm proposed in the previous subsections.

\begin{algorithm}[t]
\begin{small}
\caption{Joint Transmit Waveform and Passive Beamforming Design Algorithm}
\label{alg}
    \begin{algorithmic}[1]
    \REQUIRE $\mathbf{A}_0$, $\mathbf{B}_0$, $\mathbf{h}_\text{t}$, $\varsigma_\text{z}^2$, $P$, $\mathbf{A}_q$, $\mathbf{B}_q$, $\mathbf{h}_{\text{c},q}$, $\varsigma^2_q$, $r_q$, $\forall q$, $\mathbf{h}_{k,l}$, $\mathbf{G}_{k}$, $\gamma_{k,l}^{\text{ZF}}$, $~~~~~\gamma_{k,l}^{\text{CI}}$, $\forall k, l$, $\epsilon$, $\rho$.
    \ENSURE $\mathbf{x}^\star$, $\bm{\alpha}^\star$, and $\bm{\phi}^\star$.
        \STATE { Initialize $\bm{\phi}$, $\bm{\varphi} = \bm{\phi}$, $\mathbf{x}$, $\mathbf{y}=\mathbf{x}$, $\bm{\mu}_1 = \mathbf{0}$, and $\bm{\mu}_2 = \mathbf{0}$. }
        \WHILE {no convergence }
            \STATE{Update $\mathbf{x}$ and $\bm{\alpha}$ by solving (\ref{eq:update x}).}
            \STATE{Update $\mathbf{y}$ by (\ref{eq:ystar}).}
            \STATE{Update $\bm{\phi}$ by solving (\ref{eq:update phi}).}
            \STATE{Update $\bm{\varphi}$ by (\ref{eq:varphi star}).}
            \STATE{Update $\bm{\mu}_1$ and $\bm{\mu}_2$ by (\ref{eq:update mu}). }
        \ENDWHILE
        \STATE{Return $\mathbf{x}^\star = \mathbf{x}$, $\bm{\alpha}^\star = \bm{\alpha}$, and $\bm{\phi}^\star = \bm{\phi}$.}
    \end{algorithmic}
    \end{small}
\end{algorithm}

\subsection{Summary and Complexity Analysis}

Based on the above derivations, the joint transmit waveform and passive beamforming design for the considered RIS-aided DFRC system is straightforward and summarized in Algorithm 1.
With the initials of $\bm{\phi}$ and $\mathbf{x}$, the transmit waveform $\mathbf{x}$ and the scaling factor $\bm{\alpha}$, the auxiliary variable $\mathbf{y}$, the reflecting coefficients $\bm{\phi}$, the auxiliary variable $\bm{\varphi}$, and the dual variables $\bm{\mu}_1$ and $\bm{\mu}_2$ are alternatively updated by (\ref{eq:update x}), (\ref{eq:ystar}), (\ref{eq:update phi}), (\ref{eq:varphi star}), and (\ref{eq:update mu}), respectively, until the relative growth of the value of the AL function (\ref{eq:AL function}) is less than a preset threshold.
After obtaining the transmit waveform $\mathbf{x}^\star$, the associated optimal receive filter $\mathbf{w}^\star$ can be calculated by (\ref{eq:optimal w}).
We note that the proposed algorithm offers a locally optimal solution due to involved transformations and iterations.
Although the optimality is difficult to mathematically prove and the obtained solution only provides a lower bound of radar SINR, the effectiveness of the proposed algorithm can be verified by comparing with benchmarks in our following simulations.

Then, we briefly analyze the computational complexity of the proposed joint transmit waveform and passive beamforming design algorithm.
We assume that popular interior point method is adopted to solve the convex problems (\ref{eq:update x}), (\ref{eq:update phi}), (\ref{eq:initialize x}), and (\ref{eq:initialize x2}), whose complexity is relevant to the dimension of the variable and the number of linear matrix inequality (LMI) constraints and second-order cone (SOC) constraints.
In the initialization stage, the complexity for obtaining $\bm{\phi}$ by the Riemannian optimization is of order $\mathcal{O}\{N^{1.5}\}$.
The complexities for updating/initializing $\mathbf{x}$ under ZF-type, MMSE-type, and CI-type constraints are of order $\mathcal{O}\{\sqrt{ML+2KL}(M+K)^3L^3\}$, $\mathcal{O}\{2\sqrt{ML+KL}(M+K)^3L^3\}$, and $\mathcal{O}\{2\sqrt{ML+2KL}M^3L^3\}$, respectively.
Updating $\mathbf{y}$ by the closed-form solution (\ref{eq:ystar}) has a complexity of order $\mathcal{O}\{ML\}$.
Solving problem (\ref{eq:update phi}) to update $\bm{\phi}$ has the complexities of order $\mathcal{O}\{\sqrt{ML+2KL}N^3\}$, $\mathcal{O}\{2\sqrt{ML+KL}N^3\}$, and $\mathcal{O}\{2\sqrt{ML+2KL}N^3\}$, for the ZF-type, MMSE-type, and CI-type constraints, respectively.
The complexity of updating the closed-form $\bm{\varphi}$ by (\ref{eq:varphi star}) is of order $\mathcal{O}\{N\}$.
The complexities to update the dual variables $\bm{\mu}_1$ and $\bm{\mu}_2$ are of order $\mathcal{O}\{ML\}$ and $\mathcal{O}\{N\}$, respectively.
Therefore, the overall complexities of the proposed algorithm are of order $\mathcal{O}\{\sqrt{ML+2KL}[(M+K)^3L^3+N^3]\}$, $\mathcal{O}\{2\sqrt{ML+KL}[(M+K)^3L^3+N^3]\}$, and $\mathcal{O}\{2\sqrt{ML+2KL}(M^3L^3+N^3)\}$ for these three metrics, respectively.

\section{Simulation Studies}\label{sec:simulation results}

In this section, we provide simulation results to show the effectiveness of our proposed joint transmit waveform and passive beamforming design for RIS-aided DFRC systems.
The following settings are assumed throughout our simulations unless otherwise specified.
The BS equipped with $M = 6$ transmit/receive antennas transmits QPSK-modulated signals (i.e., $\Omega = 4$) to serve $K = 3$ downlink users and detect a target located in the range-angle position $(0,0^\circ)$, in the presence of $Q = 3$ clutter sources, which are located in $(0,-50^\circ)$, $(1,-10^\circ)$, and $(2,40^\circ)$, respectively.
The available transmit power is $P = 20$dBW and the number of waveform samples for each radar pulse is $L = 20$.
The RCS is the same for the target and the clutter sources, i.e., $\varsigma_0^2 = \varsigma_q^2 = 1,~ \forall q$.
The noise power at the radar receiver and the users are set as $\varsigma_\text{z}^2 = \sigma_k^2 = -80\text{dBm}, ~\forall k$.
The SNR requirement for the users is set as $\Gamma = \Gamma_k = 10\text{dB}, ~\forall k$.
The threshold in MMSE-type metric for assuring the MUI suppression is set as $\epsilon=10^{-9}$.

We adopt the popular path-loss model: $\text{PL}(d) = C_0(d_0/d)^\iota$, where $C_0=-30$dB, $d_0 = 1$m, $d$ represents the distance of link, and $\iota$ denotes the path-loss exponent that generally varies from 2 to 4.
In order to reap more reflection beamforming gains, the RIS is deployed in the hot-spot area near the users and the target.
We assume that the distances of the BS-target, BS-RIS, BS-user links are the same, i.e., $d_\text{t}=d_{G} = d_k = 30\text{m},~\forall k$, and the distances of the RIS-target and the RIS-user links are the same, i.e., $d_\text{rt} = d_{\text{r},k} = 3\text{m},~\forall k$.
Considering that the BS and the RIS are generally deployed at higher elevation to acquire better channel quality, we assume that the channel between the BS and RIS is LoS and stronger than the others.
Specifically, the path-loss exponents for the channels $\mathbf{G}$, $\mathbf{h}_\text{rt}$, $\mathbf{h}_{\text{rc},q}$, $\mathbf{h}_{\text{r},k}$, $\mathbf{h}_{\text{t}}$, $\mathbf{h}_{\text{c},q}$, and $\mathbf{h}_k$ are $\alpha_G = 2.5$, $\alpha_{\text{rt}} = \alpha_{\text{rc},q} = \alpha_{\text{r},k} = 2.8$, $\alpha_\text{t} = \alpha_{\text{c},q} = \alpha_k = 3, ~\forall q, k$, respectively.
In addition, the channels between the BS/RIS and the target/clutter sources are assumed to be LoS following the traditional settings in the radar field, and the channels between the BS/RIS and the users adopt the Rayleigh fading model.

\begin{figure}[t]\vspace{-0.5 cm}
\centering
\includegraphics[width = 3.4 in]{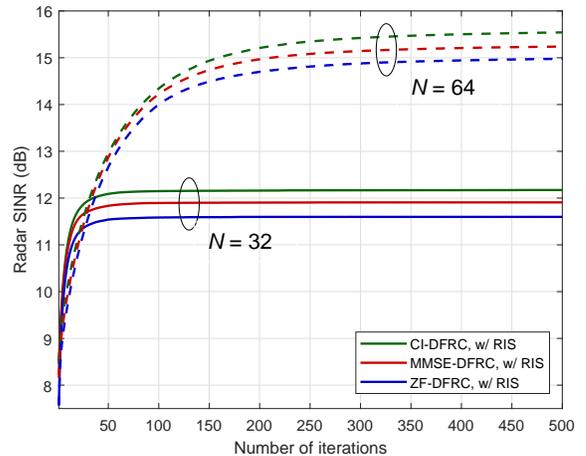}
\caption{Convergence illustration ($M = 6$, $P = 20$dBW, $\Gamma = 10$dB).}
\label{fig:iterations}\vspace{-0.2 cm}
\end{figure}

We first show the convergence performance of the proposed algorithms in Fig. \ref{fig:iterations}, where the schemes under the constraint of CI-type, MMSE-type, and ZF-type metrics are plotted and referred to as ``CI-DFRC, w/ RIS'', ``MMSE-DFRC, w/ RIS'', and ``ZF-DFRC, w/ RIS'', respectively.
The scenarios with $N = 32$ and $N = 64$ are shown with solid and dashed lines, respectively.
We observe that the radar SINRs of all schemes monotonically increase with the iterations.
Moreover, the schemes with less reflecting elements converge faster due to the lower dimensional optimizing variable.

\begin{figure}[t]
\centering
\includegraphics[width = 3.4 in]{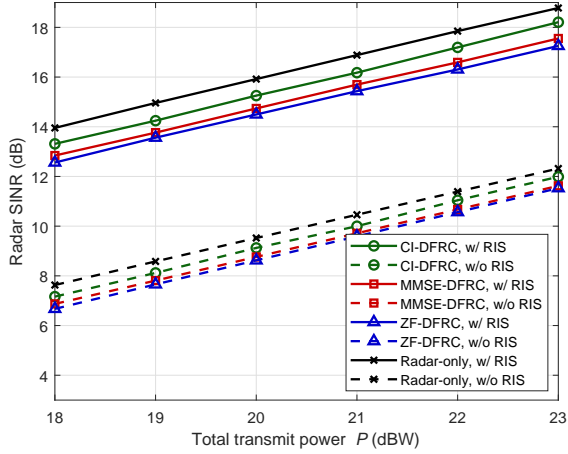}
\caption{Radar SINR versus total transmit power $P$ ($M = 6$, $N = 64$, $\Gamma = 10$dB).}
\label{fig:P}\vspace{-0.2 cm}
\end{figure}

\begin{figure}[t]
\centering
\includegraphics[width = 3.4 in]{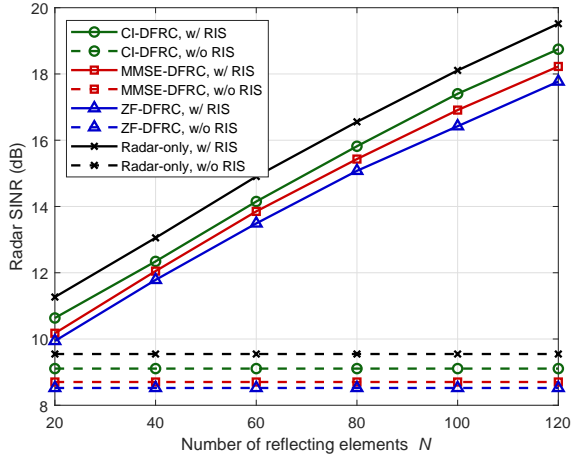}
\caption{Radar SINR versus the number of reflecting elements $N$ ($M = 6$, $P = 20$dBW, $\Gamma = 10$dB).}
\label{fig:N}\vspace{-0.2 cm}
\end{figure}

The radar SINR versus the total transmit power $P$ is presented in Fig. \ref{fig:P}.
In order to verify the effectiveness of the proposed joint transmit waveform and passive beamforming design algorithms, we also include the scenarios without RIS in the DFRC system under any one of the three proposed metrics, which are denoted as ``CI-DFRC, w/o RIS'', ``MMSE-DFRC, w/o RIS'', and ``ZF-DFRC, w/o RIS'', respectively.
The performances of a radar-only system with or without RIS are also illustrated as benchmarks, which are denoted as ``Radar-only, w/ RIS'' and ``Radar-only, w/o RIS'', respectively.
Firstly, it can be clearly observed from Fig. \ref{fig:P} that the systems with the aid of RIS (the solid lines) can achieve about 4dB radar performance improvement compared to the systems without RIS (the dash lines), which verifies the advancement of employing RIS in both DFRC and radar-only systems.
In addition, the DFRC systems can provide satisfactory communication QoS under any one of the three proposed metrics.
The DFRC systems using the CI-type communication metric achieve the highest radar SINR since it exploits the most flexibility to handle the MUI and boost the radar sensing performance, while the schemes using the ZF-type communication metric have the worst radar SINR considering that it forces the MUI to be exactly located at the direction of the desired symbol.
The performance of the scheme with the MMSE-type communication metric lies in between.
By using the CI-type metric, the DFRC system has only about 0.5dB radar performance loss compared with the radar-only system.
This phenomenon validates the effectiveness of our proposed algorithms.

Then, we illustrate the radar SINR versus the number of reflecting elements in Fig. \ref{fig:N}.
Not surprisingly, the radar SINR increases with the increasing of $N$ since more reflecting elements can provide larger passive beamforming gain to enhance both radar sensing and communication performance.
In addition, compared to the schemes without RIS, we observe that a 100-element RIS can offer about 8dB radar performance gain, since the deployment of RIS creates three more configurable propagation paths for the echo signals and enables additional DoFs for boosting the radar sensing performance.

\begin{figure}[t]
\centering
\includegraphics[width = 3.4 in]{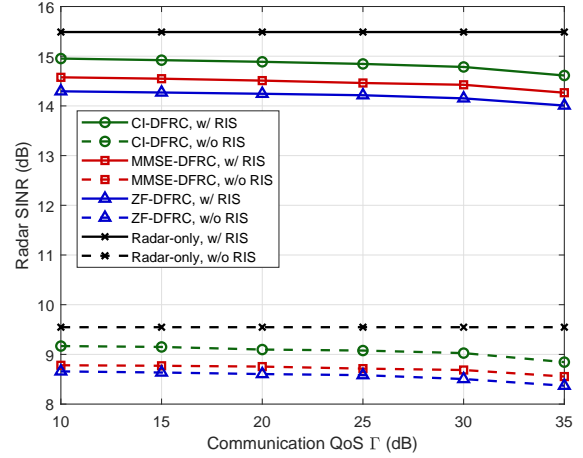}
\caption{Radar SINR versus communication QoS requirement $\Gamma$ ($M = 6$, $N = 64$, $P = 20$dBW).}
\label{fig:SNR}\vspace{-0.2 cm}
\end{figure}

\begin{figure}[t]
\centering
\includegraphics[width = 3.4 in]{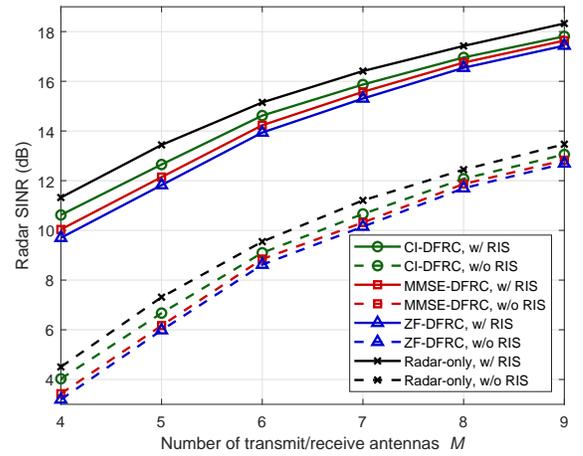}
\caption{Radar SINR versus the number of transmit/receive antennas $M$ ($N = 64$, $P = 20$dBW, $\Gamma = 10$dB).}
\label{fig:M}\vspace{-0.2 cm}
\end{figure}

Fig. \ref{fig:SNR} presents the radar SINR versus the communication QoS requirement $\Gamma$.
We can observe that the increase of $\Gamma$ has almost no influence on the radar sensing performance when $\Gamma$ is selected within a rational range, e.g., 10dB-20dB, for practical communication applications.
It is because that, when we aim to maximize radar SINR, the resulting strong transmit waveform can readily enable high quality communications and easily satisfy the practical QoS requirements using the proposed three metrics.
The trade-off between radar sensing performance and extremely high communication QoS can be seen when $\Gamma=35$dB.
The radar SINR versus the number of transmit/receive antennas is plotted in Fig. \ref{fig:M}.
Clearly, more transmit/receive antennas can exploit more spatial DoFs to improve the waveform diversity as well as to achieve higher beamforming gains.

\begin{figure}[t]
\centering
\includegraphics[width = 3.4 in]{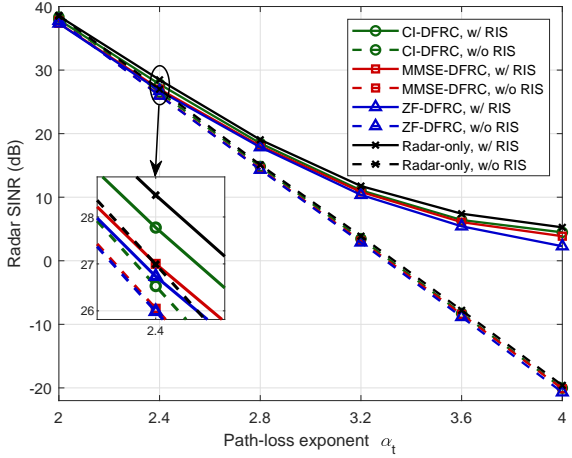}
\caption{Radar SINR versus path-loss exponent $\alpha_\text{t}$ ($M = 6$, $N = 64$, $P = 20$dBW, $\Gamma = 10$dB).}
\label{fig:alpha}\vspace{-0.2 cm}
\end{figure}

\begin{figure}[t]
\centering
\includegraphics[width = 3.4 in]{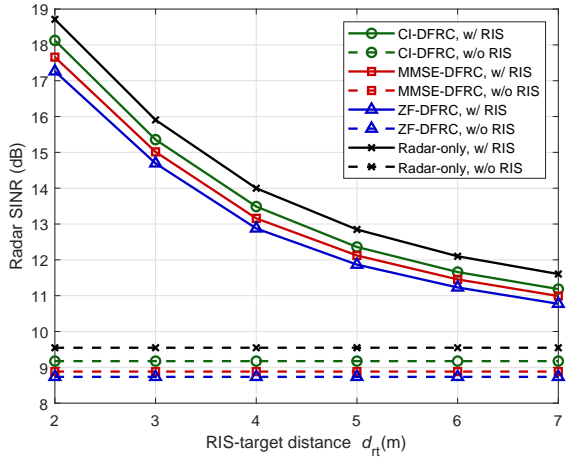}
\caption{Radar SINR versus the distance between the RIS and the target $d_\text{rt}$ ($M = 6$, $N = 64$, $P = 20$dBW, $\Gamma = 10$dB).}
\label{fig:d}\vspace{-0.2 cm}
\end{figure}

In Fig. \ref{fig:alpha}, we plot the radar SINR versus the path-loss exponent $\alpha_\text{t}$ of the direct channel between the BS and the target.
It is obvious that with the increase of $\alpha_\text{t}$, the performance gap between the schemes with and without RIS increases.
In other words, the performance improvement provided by the RIS becomes more and more pronounced as the direct channel becomes worse, or even is blocked.
For example, about 25dB gain of radar sensing can be observed when $\alpha_\text{t}=4$.
Therefore, deploying RIS in the DFRC systems with weak direct channels is a promising solution to achieve significant performance improvement; on the contrary, the impact of RIS will become negligible when the direct channel is sufficiently strong (e.g., $\alpha_\text{t}=2$).
These findings are also applicable for radar-only systems.
Finally, the radar SINR versus the distance between the RIS and the target is illustrated in Fig. \ref{fig:d}.
As excepted, higher performance improvement can be achieved when the target is closer to the RIS, which reveals that it is better to deploy the RIS near the hot-spots of interest.

\section{Conclusions}\label{sec:conclusion}

In this paper, we employed RIS in DFRC systems to achieve potential performance improvement by exploiting RIS capability of manipulating radio environment.
After modeling radar and communication signals in the considered RIS-assisted DFRC system and introducing three metrics for evaluating communication QoS, we investigated to jointly design the dual-functional transmit waveform and the passive beamforming of the RIS.
The radar output SINR was maximized subject to the constraints of communication QoS, constant-modulus transmit waveform, and RIS reflecting coefficients.
An efficient algorithm utilizing ADMM and MM methods was developed to solve the resulting complicated non-convex optimization problem.
Simulation results demonstrated the advantages of deploying RIS in DFRC systems, as well as the effectiveness of the proposed design algorithm.
Based on this initial work, we will further investigate other issues in the RIS-assisted DFRC systems, e.g., the linear block-level precoding solutions, the robustness to imperfect self-interference mitigation, the cases with imperfect or without CSI, the scenarios for multiple targets and non-zero Doppler frequencies, etc.

\begin{appendices}
\section{}
Recall that $\widetilde{\mathbf{H}}_q(\bm{\phi})\mathbf{x} = [\mathbf{I}_L\otimes \mathbf{H}_q(\bm{\phi})]\mathbf{J}_{r_q}\mathbf{x} = [\mathbf{I}_L\otimes \mathbf{A}_q+\mathbf{I}_L\otimes(\mathbf{h}_{\text{c},q}\bm{\phi}^T\mathbf{B}_q) + \mathbf{I}_L\otimes(\mathbf{B}_q^H\bm{\phi}\mathbf{h}_{\text{c},q}^H) +\mathbf{I}_L\otimes(\mathbf{B}_q^H\bm{\phi}\bm{\phi}^T\mathbf{B}_q )]\mathbf{J}_{r_q}\mathbf{x}$.
Therefore, by utilizing the property of Kronecker product \cite{Golub matrix computations}:
\begin{equation}\label{eq:Kronecker property}
\text{vec}\{\mathbf{ABC}\}=(\mathbf{C}^T\otimes\mathbf{A})\text{vec}\{\mathbf{B}\},
\end{equation}
we can extract the variable $\bm{\phi}$ from the term $\widetilde{\mathbf{H}}_q(\bm{\phi})\mathbf{x}$ based on the following transformations
\begin{subequations}\label{eq:transformations}
\begin{align}
(\mathbf{I}_L\otimes\mathbf{h}_{\text{c},q}\bm{\phi}^T\mathbf{B}_q)\mathbf{J}_{r_q}\mathbf{x}
&= \text{vec}\{\mathbf{h}_{\text{c},q}\bm{\phi}^T\mathbf{B}_q\mathbf{X}_q\mathbf{I}_L\}\\
& = [(\mathbf{B}_q\mathbf{X}_q)^T\otimes\mathbf{h}_{\text{c},q}]\bm{\phi},\\
(\mathbf{I}_L\otimes\mathbf{B}_q^H\bm{\phi}\mathbf{h}_{\text{c},q}^H)\mathbf{J}_{r_q}\mathbf{x}
&= [(\mathbf{h}_{\text{c},q}^H\mathbf{X}_q)^T\otimes\mathbf{B}_q^H]\bm{\phi},\\
(\mathbf{I}_L\otimes\mathbf{B}_q^H\bm{\phi}\bm{\phi}^T\mathbf{B}_q)\mathbf{J}_{r_q}\mathbf{x}
&= [(\mathbf{B}_q\mathbf{X}_q)^T\otimes\mathbf{B}_q^H]\text{vec}\{\bm{\phi}\bm{\phi}^T\},
\end{align}
\end{subequations}
where $\mathbf{X}_q\in\mathbb{C}^{M\times L}$ is a reshaped version of $\mathbf{J}_{r_q}\mathbf{x}$, i.e., $\mathbf{J}_{r_q}\mathbf{x} = \text{vec}\{\mathbf{X}_q\}$.
Using the results in (\ref{eq:transformations}), the term $\widetilde{\mathbf{H}}_q(\bm{\phi})\mathbf{x}$ in the objective function $f_2(\bm{\phi})$ can be equivalently and concisely re-written as
\begin{equation}\label{eq:fqx}
\widetilde{\mathbf{H}}_q(\bm{\phi})\mathbf{x} = \mathbf{a}_q + \mathbf{C}_q\bm{\phi} + \mathbf{D}_q\text{vec}\{\bm{\phi\phi}^T\},
\end{equation}
where we define
\begin{subequations}
\begin{align}
\mathbf{a}_q &\triangleq (\mathbf{I}_L\otimes\mathbf{A}_q)\mathbf{J}_{r_q}\mathbf{x},\\
\mathbf{C}_q &\triangleq (\mathbf{B}_q\mathbf{X}_q)^T\otimes\mathbf{h}_{\text{c},q} + (\mathbf{h}_{\text{c},q}^H\mathbf{X}_q)^T\otimes\mathbf{B}_q^H,\\
\mathbf{D}_q &\triangleq (\mathbf{B}_q\mathbf{X}_q)^T\otimes\mathbf{B}_q^H.
\end{align}
\end{subequations}
Based on the result in (\ref{eq:fqx}), the term $\widetilde{\mathbf{H}}_q(\bm{\phi})\mathbf{xx}^H\widetilde{\mathbf{H}}^H_q(\bm{\phi})$ in $f_2(\bm{\phi})$ can be explicitly expressed with respect to $\bm{\phi}$.
Similarly, the term $\widetilde{\mathbf{H}}_0(\bm{\phi})\mathbf{x}$ can be re-written as
\begin{equation}\label{eq:f0x}
\widetilde{\mathbf{H}}_0(\bm{\phi})\mathbf{x} = \mathbf{a}_0 + \mathbf{C}_0\bm{\phi} + \mathbf{D}_0\text{vec}\{\bm{\phi\phi}^T\},
\end{equation}
where we define
\begin{subequations}
\begin{align}
\mathbf{a}_0 &\triangleq (\mathbf{I}_L\otimes\mathbf{A}_0)\mathbf{x},\\
\mathbf{C}_0 &\triangleq (\mathbf{B}_0\mathbf{X}_0)^T\otimes\mathbf{h}_{\text{t}} + (\mathbf{h}_{\text{t}}^H\mathbf{X}_0)^T\otimes\mathbf{B}_0^H,\\
\mathbf{D}_0 &\triangleq (\mathbf{B}_0\mathbf{X}_0)^T\otimes\mathbf{B}_0^H,
\end{align}
\end{subequations}
and $\mathbf{X}_0\in\mathbb{C}^{M\times L}$ is a reshaped version of $\mathbf{x}$, i.e., $\mathbf{x} = \text{vec}\{\mathbf{X}_0\}$.
Plugging the results in (\ref{eq:fqx}) and (\ref{eq:f0x}) into (\ref{eq:fphi}), the objective function $f_2(\bm{\phi})$ can be concisely re-formulated as
\begin{equation}\label{eq:reform f2}
\begin{aligned}
f_2(\bm{\phi}) & = \bm{\phi}^H\mathbf{F}_t\bm{\phi}+\Re\{\bm{\phi}^H\mathbf{f}_t\}
+\mathbf{v}^H\mathbf{F}_{\text{v},t}\mathbf{v}+\Re\{\mathbf{v}^H\mathbf{f}_{\text{v},t}\}\\
&\quad+\Re\{\mathbf{v}^H\mathbf{L}_t\bm{\phi}\} + c_2 + \frac{\rho}{2}\big\|\bm{\phi}-\bm{\varphi}_t+\frac{\bm{\mu}_2}{\rho}\big\|^2,
\end{aligned}\end{equation}
where for conciseness we define
\begin{subequations}\label{eq:fl definition}
\begin{align}
\mathbf{v} &\triangleq \text{vec}\{\bm{\phi\phi}^T\},\\
\mathbf{F}_t &\triangleq \sum_{q=1}^Q\varsigma_q^2\mathbf{C}_q^H\mathbf{M}_t^{-1}
\mathbf{s}_t\mathbf{s}_t^H\mathbf{M}_t^{-1}\mathbf{C}_q, \\
\mathbf{f}_t &\triangleq 2\sum_{q=1}^Q\varsigma_q^2\mathbf{C}_q^H\mathbf{M}_t^{-1}
\mathbf{s}_t\mathbf{s}_t^H\mathbf{M}_t^{-1}\mathbf{a}_q - 2\mathbf{C}_0^H\mathbf{M}_t^{-1}\mathbf{s}_t, \\
\mathbf{F}_{\text{v},t}&\triangleq \sum_{q=1}^Q\varsigma_q^2\mathbf{D}_q^H\mathbf{M}_t^{-1}
\mathbf{s}_t\mathbf{s}_t^H\mathbf{M}_t^{-1}\mathbf{D}_q, \\
\mathbf{f}_{\text{v},t}&\triangleq 2\sum_{q=1}^Q\varsigma_q^2\mathbf{D}_q^H\mathbf{M}_t^{-1}
\mathbf{s}_t\mathbf{s}_t^H\mathbf{M}_t^{-1}\mathbf{a}_q - 2\mathbf{D}_0^H\mathbf{M}_t^{-1}\mathbf{s}_t,\\
\mathbf{L}_t &\triangleq 2\sum_{q=1}^Q\varsigma_q^2\mathbf{D}_q^H\mathbf{M}_t^{-1}
\mathbf{s}_t\mathbf{s}_t^H\mathbf{M}_t^{-1}\mathbf{C}_q,\\
c_2 &\triangleq \sum_{q=1}^Q\varsigma_q^2\mathbf{s}_t^H\mathbf{M}_t^{-1}\mathbf{a}_q\mathbf{a}_q^H\mathbf{M}_t^{-1}\mathbf{s}_t - 2\mathbf{s}_t^H\mathbf{M}_t^{-1}\mathbf{a}_0.
\end{align}
\end{subequations}

Furthermore, considering the significantly high cost of computing and storing the $N^2\times L$-dimensional matrix $\mathbf{C}_q$ and $ML\times N^2$-dimensional matrix $\mathbf{D}_q$, we propose to re-write the expressions in (\ref{eq:fl definition}) to eliminate the intermediate variables $\mathbf{C}_q$ and $\mathbf{D}_q,~\forall q = 0, 1, \ldots, Q$, to reduce the computational complexity for the following derivations.
Specifically, by utilizing the transformation in (\ref{eq:Kronecker property}), we have
\begin{subequations}\label{eq:Dqst}
\begin{align}
\mathbf{D}_q^H\mathbf{M}_t^{-1}\mathbf{s}_t& = [(\mathbf{X}_q^H\mathbf{B}_q^H)^T\otimes\mathbf{B}_q]\mathbf{M}_t^{-1}\mathbf{s}_t\\
& = \text{vec}\big\{\mathbf{B}_q\widetilde{\mathbf{M}}_t\mathbf{X}_q^H\mathbf{B}_q^H\big\},\\
\mathbf{C}_q^H\mathbf{M}_t^{-1}\mathbf{s}_t &= [(\mathbf{X}_q^H\mathbf{B}_q^H)^T\hspace{-0.05 cm}\otimes\hspace{-0.05 cm}\mathbf{h}_{\text{c},q}^H \hspace{-0.05 cm}+ \hspace{-0.05 cm}(\mathbf{X}_q^H\mathbf{h}_{\text{c},q})^T\hspace{-0.05 cm}\otimes\hspace{-0.05 cm}\mathbf{B}_q]\mathbf{M}_t^{-1}\mathbf{s}_t\\
& = \text{vec}\big\{\mathbf{h}_{\text{c},q}^H\widetilde{\mathbf{M}}_t\mathbf{X}_q^H\mathbf{B}_q^H
+ \mathbf{B}_q\widetilde{\mathbf{M}}_t\mathbf{X}_q^H\mathbf{h}_{\text{c},q}\big\}\\
& = \mathbf{B}_q^*\mathbf{X}_q^*\widetilde{\mathbf{M}}_t^T\mathbf{h}_{\text{c},q}^* +
\mathbf{B}_q\widetilde{\mathbf{M}}_t\mathbf{X}_q^H\mathbf{h}_{\text{c},q},
\end{align}
\end{subequations}
where $\widetilde{\mathbf{M}}_t\in\mathbb{C}^{M\times L}$ is a reshaped version of $\mathbf{M}_t^{-1}\mathbf{s}_t$, i.e., $\mathbf{M}_t^{-1}\mathbf{s}_t = \text{vec}\{\widetilde{\mathbf{M}}_t\}$.
Therefore, by defining
\begin{subequations}\label{eq:Dqcq}
\begin{align}
\widetilde{\mathbf{D}}_{q,t}&\triangleq\mathbf{B}_q\widetilde{\mathbf{M}}_t\mathbf{X}_q^H\mathbf{B}_q^H,\\
\widetilde{\mathbf{c}}_{q,t}&\triangleq \mathbf{B}_q^*\mathbf{X}_q^*\widetilde{\mathbf{M}}_t^T\mathbf{h}_{\text{c},q}^* +
\mathbf{B}_q\widetilde{\mathbf{M}}_t\mathbf{X}_q^H\mathbf{h}_{\text{c},q},
\end{align}
\end{subequations}
we can express $\mathbf{F}_t$, $\mathbf{f}_t$, $\mathbf{F}_{\text{v},t}$, $\mathbf{f}_{\text{v},t}$, and $\mathbf{L}_t$ in concise terms as
\begin{subequations}\label{eq:fl reform}
\begin{align}
\mathbf{F}_t &= \sum_{q=1}^Q\varsigma_q^2\widetilde{\mathbf{c}}_{q,t}\widetilde{\mathbf{c}}_{q,t}^H,\\
\mathbf{f}_t&= 2\sum_{q=1}^Q\varsigma_q^2\mathbf{s}_t^H\mathbf{M}_t^{-1}\mathbf{a}_q\widetilde{\mathbf{c}}_{q,t} - 2\widetilde{\mathbf{c}}_{0,t},\\
\mathbf{F}_{\text{v},t}&= \sum_{q=1}^Q\varsigma_q^2\text{vec}\{\widetilde{\mathbf{D}}_{q,t}\}\text{vec}^H\{\widetilde{\mathbf{D}}_{q,t}\},\\
\mathbf{f}_{\text{v},t}&= 2\sum_{q=1}^Q\varsigma_q^2\mathbf{s}_t^H\mathbf{M}_t^{-1}\mathbf{a}_q\text{vec}\{\widetilde{\mathbf{D}}_{q,t}\}
-2\text{vec}\{\widetilde{\mathbf{D}}_{0,t}\},\\
\mathbf{L}_t &= 2\sum_{q=1}^Q\varsigma_q^2\text{vec}\{\widetilde{\mathbf{D}}_{q,t}\}\widetilde{\mathbf{c}}_{q,t}^H.
\end{align}
\end{subequations}

Then, we also use the property of Kronecker product (\ref{eq:Kronecker property}) to handle the communication QoS constraints (\ref{eq:communication QoS constraints}\text{a})-(\ref{eq:communication QoS constraints}\text{c}).
Specifically, we have the following transformation
\begin{subequations}\begin{align}
[\mathbf{h}^H_{k,l}+(\mathbf{e}_l^T\otimes\bm{\phi}^T)\mathbf{G}_{k}]\mathbf{x}
&= \mathbf{h}^H_{k,l}\mathbf{x} + \bm{\phi}^T\widetilde{\mathbf{G}}_{k}\mathbf{e}_l,\\
& = \mathbf{h}^H_{k,l}\mathbf{x} + \widetilde{\mathbf{g}}_{k,l}^T\bm{\phi},
\end{align}\end{subequations}
where $\widetilde{\mathbf{G}}_{k}\in\mathbb{C}^{N\times L}$ is a reshaped version of $\mathbf{G}_{k}\mathbf{x}$, i.e., $\mathbf{G}_{k}\mathbf{x} = \text{vec}\{\widetilde{\mathbf{G}}_{k}\}$, and we define $\widetilde{\mathbf{g}}_{k,l}\triangleq\widetilde{\mathbf{G}}_{k}\mathbf{e}_l$ for simplicity.
Thus, the communication QoS constraints (\ref{eq:communication QoS constraints}\text{a})-(\ref{eq:communication QoS constraints}\text{c}) can be concisely re-written as
\begin{subequations}\label{eq:reform QoS}
\begin{align}
&\mathbf{h}^H_{k,l}\mathbf{x} + \widetilde{\mathbf{g}}_{k,l}^T\bm{\phi} =\alpha_{k,l}\gamma_{k,l}^\text{ZF},~~\alpha_{k,l} \geq 1,~~\forall k,l, \\
&\big|\mathbf{h}^H_{k,l}\mathbf{x}\hspace{-0.05 cm}+\hspace{-0.05 cm}\widetilde{\mathbf{g}}_{k,l}^T\bm{\phi} \hspace{-0.05 cm}-\hspace{-0.05 cm}\alpha_{k,l}\gamma_{k,l}^\text{ZF}\big|^2\hspace{-0.05 cm}\leq\hspace{-0.05 cm}\epsilon,~\alpha_{k,l}\hspace{-0.05 cm}\geq\hspace{-0.05 cm}1,~\forall k,l,\\
&\Re\big\{\gamma_{k,l}^\text{CI}(\mathbf{h}^H_{k,l}\mathbf{x} + \widetilde{\mathbf{g}}_{k,l}^T\bm{\phi})\big\} \geq 1,~~\forall k,l.
\end{align}
\end{subequations}

Combining the re-formulations of the objective function $f_2(\bm{\phi})$ in (\ref{eq:reform f2}) and the communication QoS constraints (\ref{eq:reform QoS}), problem (\ref{eq:problem phi}) can be equivalently expressed as problem (\ref{eq:problem phi reform}).

\section{}

Using the re-formulated version of $\mathbf{F}_{\text{v},t}$ in (\ref{eq:fl reform}c), we have
\begin{subequations}\label{eq:Fv}
\begin{align}
\mathbf{F}_{\text{v},t}\mathbf{v}_t &=\sum_{q=1}^Q\varsigma_q^2\text{vec}\{\widetilde{\mathbf{D}}_{q,t}\}\text{vec}^H\{\widetilde{\mathbf{D}}_{q,t}\}
\text{vec}\{\bm{\phi}\bm{\phi}^T\}\\
& = \sum_{q=1}^Q\varsigma_q^2\text{vec}\{\widetilde{\mathbf{D}}_{q,t}\}\text{Tr}\{\bm{\phi}_t\bm{\phi}_t^T\widetilde{\mathbf{D}}_{q,t}^H\}\\
& = \sum_{q=1}^Q\varsigma_q^2\bm{\phi}_t^T\widetilde{\mathbf{D}}_{q,t}^H\bm{\phi}_t\text{vec}\{\widetilde{\mathbf{D}}_{q,t}\},
\end{align}
\end{subequations}
where (\ref{eq:Fv}b) holds by applying the transformation \cite{Golub matrix computations}:
\be
\text{Tr}\{\mathbf{ABCD}\} = \text{vec}^H\{\mathbf{D}^H\}(\mathbf{C}^T\otimes\mathbf{A})\text{vec}\{\mathbf{B}\}.
\ee
Plugging the result in (\ref{eq:Fv}c) into (\ref{eq:fvttilde}), the vector $\widetilde{\mathbf{f}}_{\text{v},t}$ can be re-written as
\be\label{eq:fvttlr}
\widetilde{\mathbf{f}}_{\text{v},t}= 2\sum_{q=1}^Q\varsigma_q^2\bm{\phi}_t^T\widetilde{\mathbf{D}}_{q,t}^H\bm{\phi}_t\text{vec}\{\widetilde{\mathbf{D}}_{q,t}\}
-2\lambda_1\text{vec}\{\bm{\phi}\bm{\phi}^T\}.
\ee
Combining with the re-formulated $\mathbf{f}_{\text{v},t}$ in (\ref{eq:fl reform}d), the matrix version of $\widetilde{\mathbf{f}}_{\text{v},t}+\mathbf{f}_{\text{v},t}$, i.e., $\widetilde{\mathbf{F}}_{\text{v},t}$, can be expressed as
\be
\widetilde{\mathbf{F}}_{\text{v},t}\hspace{-0.06cm}= \hspace{-0.06cm} 2\hspace{-0.08cm}\sum_{q=1}^Q\hspace{-0.08cm}\varsigma_q^2(\bm{\phi}_t^T\widetilde{\mathbf{D}}_{q,t}^H\bm{\phi}_t\hspace{-0.03cm}
+\hspace{-0.03cm}\mathbf{s}_t^H\mathbf{M}_t^{-1}\mathbf{a}_q)\widetilde{\mathbf{D}}_{q,t}
\hspace{-0.03cm}-\hspace{-0.03cm}2\widetilde{\mathbf{D}}_{0,t}\hspace{-0.03cm}-\hspace{-0.03cm}2\lambda_1\bm{\phi}_t\bm{\phi}_t^T.
\ee

\end{appendices}

\end{document}